\documentstyle[12pt]{art12}
\makeatletter
\let\reset@font\empty

\def\indexname{Index}
\def\figurename{Figure}
\def\tablename{Table}
\def\abstractname{Abstract}

\def\@ptsize{0}
\@namedef{ds@11pt}{\def\@ptsize{1}}
\@namedef{ds@12pt}{\def\@ptsize{2}}
\def\ds@twoside{\@twosidetrue
           \@mparswitchtrue}

\def\ds@draft{\overfullrule 5\p@}

\newif\if@titlepage \@titlepagefalse
\def\ds@titlepage{\@titlepagetrue}

\def\ds@twocolumn{\@twocolumntrue}

\newdimen\mathindent
\newif\ifletter
\newif\ifpmb
\newlength{\varind}
\newlength{\figdepth}
\newlength{\figwidth}
\newlength{\secfigwidth}

\newlength{\indentedwidth}

\newcounter{jnl}
\newcounter{yr}
\newcounter{tabtype}
\newcounter{figtype}
\newcounter{eqnval}
\@twosidetrue
\def\ds@draft{\overfullrule 5\p@}
\@options

\def\@normalsize{\@setsize\normalsize{16pt}\xiipt\@xiipt
  \abovedisplayskip 12pt plus3pt minus6pt
  \belowdisplayskip \abovedisplayskip
  \abovedisplayshortskip \z@ plus4pt
  \belowdisplayshortskip 7pt plus4pt minus4pt}
\def\small{\@setsize\small{14pt}\xipt\@xipt
  \abovedisplayskip 10pt plus 3pt minus 4pt
  \belowdisplayskip \abovedisplayskip
  \abovedisplayshortskip \z@ plus3pt
  \belowdisplayshortskip 5pt plus3pt minus 3pt
  \def\@listi{\topsep 5pt plus 3pt minus 3pt\parsep 0pt plus 1pt
         \itemsep \parsep}}
\def\footnotesize{\@setsize\footnotesize{14pt}\xpt\@xpt
  \abovedisplayskip 7pt plus 3pt minus 4pt
  \belowdisplayskip \abovedisplayskip
  \abovedisplayshortskip \z@ plus 2pt
  \belowdisplayshortskip 3pt plus 1pt minus2pt
  \def\@listi{\topsep 4pt plus 2pt minus 2pt\parsep 0pt plus 1pt
         \itemsep \parsep}}
\def\scriptsize{\@setsize\scriptsize{13pt}\ixpt\@ixpt}
\def\tiny{\@setsize\tiny{10pt}\viipt\@viipt}
\def\large{\@setsize\large{18pt}\xivpt\@xivpt}
\def\Large{\@setsize\Large{22pt}\xviipt\@xviipt}
\def\LARGE{\@setsize\LARGE{25pt}\xxpt\@xxpt}
\def\huge{\@setsize\huge{30pt}\xxvpt\@xxvpt}
\def\Huge{\@setsize\Huge{30pt}\xxvpt\@xxvpt}
\normalsize
\skip\footins 12\p@ plus 4\p@ minus 2\p@
\@beginparpenalty -\@lowpenalty
\@endparpenalty -\@lowpenalty
\@itempenalty -\@lowpenalty

\@noskipsecfalse   

\def\section{\@startsection{section}{1}{\z@}{-3.5ex plus -1ex minus
 -.2ex}{2.3ex plus .2ex}{\noindent\reset@font\normalsize\bf\raggedright}}
\def\subsection{\@startsection{subsection}{2}{\z@}{-3.25ex plus -1ex minus
 -.2ex}{1.5ex plus .2ex}{\noindent\reset@font
  \normalsize\it\raggedright\nohyphens}}
\def\subsubsection{\@startsection{subsubsection}{3}{\z@}{-3.25ex plus
-1ex minus -.2ex}{-1em}{\reset@font\normalsize\it\nohyphens}}
\def\paragraph{\@startsection
 {paragraph}{4}{\z@}{3.25ex plus 1ex minus
.2ex}{-1em}{\reset@font\normalsize\it}}
\def\subparagraph{\@startsection
 {subparagraph}{4}{\parindent}{3.25ex plus 1ex minus
 .2ex}{-1em}{\reset@font\normalsize\it}}

\def\@sect#1#2#3#4#5#6[#7]#8{\ifnum #2>\c@secnumdepth
     \let\@svsec\@empty\else
     \refstepcounter{#1}\edef\@svsec{\csname the#1\endcsname.\hskip 1em}\fi
     \@tempskipa #5\relax
      \ifdim \@tempskipa>\z@
        \begingroup #6\relax
          \noindent{\hskip #3\relax\@svsec}{\interlinepenalty \@M #8\par}%
        \endgroup
       \csname #1mark\endcsname{#7}\addcontentsline
         {toc}{#1}{\ifnum #2>\c@secnumdepth \else
                      \protect\numberline{\csname the#1\endcsname}\fi
                    #7}\else
        \def\@svsechd{#6\hskip #3\relax  
                   \@svsec #8\csname #1mark\endcsname
                      {#7}\addcontentsline
                           {toc}{#1}{\ifnum #2>\c@secnumdepth \else
                             \protect\numberline{\csname the#1\endcsname}\fi
                       #7}}\fi
     \@xsect{#5}}
\def\@ssect#1#2#3#4#5{\@tempskipa #3\relax
   \ifdim \@tempskipa>\z@
     \begingroup #4\noindent{\hskip #1}{\interlinepenalty
   \@M #5\par}\endgroup
   \else \def\@svsechd{#4\hskip #1\relax #5}\fi
    \@xsect{#3}}

\def\appendix{\@@par
 \setcounter{section}{0}
 \setcounter{subsection}{0}
 \setcounter{subsubsection}{0}
 \setcounter{equation}{0}
 \setcounter{figure}{0}
 \setcounter{table}{0}
 \def\thesection{Appendix \Alph{section}}
 \def\theequation{\ifnumbysec
      \Alph{section}.\arabic{equation}\else
      \Alph{section}\arabic{equation}\fi}
 \def\thetable{\ifnumbysec
      \Alph{section}\arabic{table}\else
      A\arabic{table}\fi}
 \def\thefigure{\ifnumbysec
      \Alph{section}\arabic{figure}\else
      A\arabic{figure}\fi}}

\leftmargin\leftmargini
\labelwidth\leftmargini\advance\labelwidth-\labelsep
\def\@listI{\leftmargin\leftmargini \parsep 4\p@ plus2\p@ minus\p@
\topsep 8\p@ plus2\p@ minus4\p@
\itemsep 4\p@ plus2\p@ minus\p@}

\let\@listi\@listI
\@listi

\def\@listii{\leftmargin\leftmarginii
 \labelwidth\leftmarginii\advance\labelwidth-\labelsep
 \topsep 3\p@ plus 1\p@ minus 1\p@
 \parsep 0\p@ plus 1\p@
 \itemsep \parsep}
\def\@listiii{\leftmargin\leftmarginiii
 \labelwidth\leftmarginiii\advance\labelwidth-\labelsep
 \topsep 2\p@ plus 1\p@ minus 1\p@
 \parsep \z@ \partopsep 1\p@ plus 0\p@ minus 1\p@
 \itemsep \topsep}
\def\@listiv{\leftmargin\leftmarginiv
 \labelwidth\leftmarginiv\advance\labelwidth-\labelsep}
\def\@listv{\leftmargin\leftmarginv
 \labelwidth\leftmarginv\advance\labelwidth-\labelsep}
\def\@listvi{\leftmargin\leftmarginvi
 \labelwidth\leftmarginvi\advance\labelwidth-\labelsep}

\def\hexnumber@#1{\ifcase#1 0\or 1\or 2\or 3\or 4\or 5\or 6\or 7\or 8\or
 9\or A\or B\or C\or D\or E\or F\fi}
\edef\bffam@{\hexnumber@\bffam}
\mathchardef\bGamma "0\bffam@00
\mathchardef\bDelta "0\bffam@01
\mathchardef\bTheta "0\bffam@02
\mathchardef\bLambda "0\bffam@03
\mathchardef\bXi "0\bffam@04
\mathchardef\bPi "0\bffam@05
\mathchardef\bSigma "0\bffam@06
\mathchardef\bUpsilon "0\bffam@07
\mathchardef\bPhi "0\bffam@08
\mathchardef\bPsi "0\bffam@09
\mathchardef\bOmega "0\bffam@0A

\def\theenumi{\roman{enumi}}

\def\theenumii{\alph{enumii}}
\def\p@enumii{\theenumi.}

\def\theenumiii{\arabic{enumiii}}
\def\p@enumiii{\p@enumii.\theenumii}

\def\p@enumiv{\p@enumiii.\theenumiii}

\def\labelitemi{$\m@th\bullet$}

\def\labelitemiii{$\m@th\ast$}
\def\labelitemiv{$\m@th\cdot$}

\def\verse{\let\\=\@centercr
 \list{}{\itemsep\z@ \itemindent -1.5em\listparindent \itemindent
 \rightmargin\leftmargin\advance\leftmargin 1.5em}\item[]}

\def\quotation{\list{}{\listparindent 1.5em
 \itemindent\listparindent
 \rightmargin\leftmargin\parsep 0\p@ plus 1\p@}\item[]}

\def\descriptionlabel#1{\hspace\labelsep \bf #1}
\def\description{\list{}{\labelwidth\z@ \itemindent-\leftmargin
 \let\makelabel\descriptionlabel}}

\def\enumerate{\ifnum \@enumdepth >3 \@toodeep\else
      \advance\@enumdepth \@ne
      \edef\@enumctr{enum\romannumeral\the\@enumdepth}\list
      {\csname label\@enumctr\endcsname}{\usecounter
        {\@enumctr}\def\makelabel##1{##1\hss}}\fi}
\def\itemize{\ifnum \@itemdepth >3 \@toodeep\else \advance\@itemdepth \@ne
\edef\@itemitem{labelitem\romannumeral\the\@itemdepth}%
\list{\csname\@itemitem\endcsname}{\def\makelabel##1{##1\hss}\topsep=3pt
  \parsep=0pt\listparindent=0pt\itemsep=0pt\partopsep=0pt\rightmargin=0pt
  }\fi}
\tabbingsep \labelsep
\skip\@mpfootins = \skip\footins
\def\titlepage{\@restonecolfalse\if@twocolumn\@restonecoltrue\onecolumn
     \else \newpage \fi \thispagestyle{myheadings}\c@page\z@}

\def\endtitlepage{\if@restonecol\twocolumn \else \newpage \fi}

\newcounter {section}
\newcounter {subsection}[section]
\newcounter {subsubsection}[subsection]
\newcounter {paragraph}[subsubsection]
\newcounter {subparagraph}[paragraph]

\def\thesection {\arabic{section}}

\def\@chapapp{Section}

\def\@pnumwidth{1.55em}
\def\@tocrmarg {2.55em}
\def\@dotsep{4.5}
\def\tableofcontents{\@restonecolfalse\if@twocolumn\@restonecoltrue
 \onecolumn\fi\section*{Contents}{}\thispagestyle{empty}
 \@starttoc{toc}\if@restonecol\twocolumn\fi}
\def\l@section{\@dottedtocline{1}{1.5em}{2.3em}}
\def\l@subsection{\@dottedtocline{2}{3.8em}{3.2em}}
\def\l@subsubsection{\@dottedtocline{3}{7.0em}{4.1em}}
\def\l@paragraph{\@dottedtocline{4}{10em}{5em}}
\def\l@subparagraph{\@dottedtocline{5}{12em}{6em}}
\def\listoffigures{\@restonecolfalse\if@twocolumn\@restonecoltrue\onecolumn
 \fi\section*{List of Figures\@mkboth
 {LIST OF FIGURES}{LIST OF FIGURES}}\@starttoc{lof}\if@restonecol\twocolumn
 \fi}
\def\l@figure{\@dottedtocline{1}{1.5em}{2.3em}}
\def\listoftables{\@restonecolfalse\if@twocolumn\@restonecoltrue\onecolumn
 \fi\section*{List of Tables\@mkboth
 {LIST OF TABLES}{LIST OF TABLES}}\@starttoc{lot}\if@restonecol\twocolumn
 \fi}
\let\l@table\l@figure
\def\@dottedtocline#1#2#3#4#5{\ifnum #1>\c@tocdepth \else
  \vskip \z@ plus .2\p@
  {\leftskip #2\relax \rightskip \@tocrmarg \parfillskip -\rightskip
    \parindent #2\relax\@afterindenttrue
   \interlinepenalty\@M
   \leavevmode
   \@tempdima #3\relax \advance\leftskip \@tempdima
   \hbox{}\hskip -\leftskip
    #4\nobreak\hfill \nobreak \hbox to\@pnumwidth{\hfil
   \rm #5}\@@par}\fi}

\@addtoreset{footnote}{page}
\long\def\@makefntext#1{\parindent 1em\noindent
 \makebox[1em][l]{\footnotesize\rm$\m@th{\fnsymbol{footnote}}$}%
 \footnotesize\rm #1}
\def\@makefnmark{\hbox{${\fnsymbol{footnote}}\m@th$}}
\def\@thefnmark{\fnsymbol{footnote}}
\def\footnote{\@ifnextchar[{\@xfootnote}{\stepcounter{\@mpfn}%
       \begingroup\let\protect\noexpand
       \xdef\@thefnmark{\thempfn}\endgroup
     \@footnotemark\@footnotetext}}
\def\@fnsymbol#1{\ifcase#1\or \dagger\or \ddagger\or \S\or
   \|\or \P\or ^{+}\or ^{\tsty *}\or \sharp
   \or \dagger\dagger \else\@ctrerr\fi\relax}

\def\[{\relax\ifmmode\@badmath\else
 \begin{trivlist}
 \@beginparpenalty\predisplaypenalty
 \@endparpenalty\postdisplaypenalty
 \item[]\leavevmode
 \hbox to\linewidth\bgroup$ \displaystyle
 \hskip\mathindent\bgroup\fi}
\def\]{\relax\ifmmode \egroup $\hfil \egroup \end{trivlist}\else \@badmath \fi}
\def\equation{\@beginparpenalty\predisplaypenalty
 \@endparpenalty\postdisplaypenalty
\refstepcounter{equation}\trivlist \item[]\leavevmode
 \hbox to\linewidth\bgroup $ \displaystyle
\hskip\mathindent}
\def\endequation{$\hfil \displaywidth\linewidth\@eqnnum\egroup \endtrivlist}
\def\eqnarray{\stepcounter{equation}\let\@currentlabel=\theequation
\global\@eqnswtrue
\global\@eqcnt\z@\tabskip\mathindent\let\\=\@eqncr
\abovedisplayskip\topsep\ifvmode\advance\abovedisplayskip\partopsep\fi
\belowdisplayskip\abovedisplayskip
\belowdisplayshortskip\abovedisplayskip
\abovedisplayshortskip\abovedisplayskip
$$\halign to
\linewidth\bgroup\@eqnsel$\displaystyle\tabskip\z@
 {##{}}$&\global\@eqcnt\@ne $\displaystyle{{}##{}}$\hfil    
 &\global\@eqcnt\tw@ $\displaystyle{{}##}$\hfil
 \tabskip\@centering&\llap{##}\tabskip\z@\cr}
\def\endeqnarray{\@@eqncr\egroup
 \global\advance\c@equation\m@ne$$\global\@ignoretrue }
\newcommand{\jl}[1]{\setcounter{jnl}{#1}%
    \ifnum\thejnl=12\global\pmbtrue\fi
    \ifnum\thejnl=15\global\pmbtrue\fi}
\def\journal{\ifnum\thejnl=1 J. Phys.\ A: Math.\ Gen.\
        \else\ifnum\thejnl=2 J. Phys.\ B: At.\ Mol.\ Opt.\ Phys.\
        \else\ifnum\thejnl=3 J. Phys.:\ Condens. Matter\
        \else\ifnum\thejnl=4 J. Phys.\ G: Nucl.\ Part.\ Phys.\
        \else\ifnum\thejnl=5 Inverse Problems\
        \else\ifnum\thejnl=6 Class. Quantum Grav.\
        \else\ifnum\thejnl=7 Network\
        \else\ifnum\thejnl=8 Nonlinearity\
        \else\ifnum\thejnl=9 Quantum Opt.\
        \else\ifnum\thejnl=10 Waves in Random Media\
        \else\ifnum\thejnl=11 Pure Appl. Opt.\
        \else\ifnum\thejnl=12 Phys. Med. Biol.\ %
        \else\ifnum\thejnl=13 Modelling Simul.\ Mater.\ Sci.\ Eng.\
        \else\ifnum\thejnl=14 Plasma Phys. Control. Fusion\
        \else\ifnum\thejnl=15 Physiol. Meas.\
        \else\ifnum\thejnl=16 Sov.\ Lightwave Commun.\
        \else\ifnum\thejnl=17 High Perform.\ Polym.\
        \else\ifnum\thejnl=18 J.\ Hard Mater.\
        \else\ifnum\thejnl=19 J.\ Phys.\ D: Appl.\ Phys.\
        \else\ifnum\thejnl=20 Supercond.\ Sci.\ Technol.\
        \else\ifnum\thejnl=21 Semicond.\ Sci.\ Technol.\
        \else\ifnum\thejnl=22 Nanotechnology\
        \else\ifnum\thejnl=23 Meas.\ Sci.\ Technol.\
        \else\ifnum\thejnl=24 Plasma Source Sci.\ Technol.\
        \else\ifnum\thejnl=25 Smart Mater.\ Struct.\
        \else\ifnum\thejnl=26 J.\ Micromech.\ Microeng.\
        \else\ifnum\thejnl=27 Distrib.\ Syst.\ Engng\
\else Institute of Physics Publishing
\fi\fi\fi\fi\fi\fi\fi\fi\fi\fi\fi\fi\fi\fi\fi
\fi\fi\fi\fi\fi\fi\fi\fi\fi\fi\fi\fi}

\def\catchline{\hfill}

\def\cpyrtline{\hfill}
\def\maketitle{\vspace*{\baselineskip}\vspace{0\p@ plus1fil}
    \noindent Short title: \@shorttitle\par
    \@submitted
    \vspace*{\baselineskip}
    \noindent\today\par\newpage}
\def\@rticle#1#2{\thispagestyle{myheadings}%
     \vspace*{.5pc}%
    {\parindent=\mathindent \bf #1\par}%
     \vspace*{1.5pc}%
    {\exhyphenpenalty=10000\hyphenpenalty=10000
     \Large\raggedright\noindent
     \bf#2\par}\def\@shorttitle{#1}\futurelet\next\sh@rttitle}%
\def\title#1{\def\@shorttitle{#1}%
    \thispagestyle{myheadings}%
    \vspace*{3pc}{\exhyphenpenalty=10000\hyphenpenalty=10000
    \Large\raggedright\noindent
    \bf#1\par}\futurelet\next\sh@rttitle}

\def\article#1#2{\@rticle{#1}{#2}}
\def\review#1{\@rticle{REVIEW \ifpmb\else ARTICLE\fi}{#1}}
\def\topical#1{\@rticle{TOPICAL REVIEW}{#1}}
\def\ireview#1{\@rticle{INTRODUCTORY REVIEW}{#1}}
\def\comment#1{\@rticle{COMMENT}{#1}}
\def\note#1{\@rticle{NOTE}{#1}}
\def\prelim#1{\@rticle{PRELIMINARY COMMUNICATION}{#1}}
\def\letter#1{\@rticle{LETTER TO THE EDITOR}{#1}}
\def\sh@rttitle{\ifx\next[\let\next=\sh@rt
                \else\let\next=\f@ll\fi\next}
\def\sh@rt[#1]{\gdef\@shorttitle{#1}}
\def\f@ll{}
\renewcommand{\author}[1]{\vspace*{1.5pc}%
   \begin{indented}%
   \item[]\normalsize\ifnum\thejnl=8\bf\else\rm\fi\raggedright#1
   \end{indented}%
   \smallskip}
\def\abstract{\vspace{16pt plus3pt minus3pt}
   \begin{indented}
   \item[]{\bf \abstractname.}\quad\rm\ignorespaces}%
\def\endabstract{\end{indented}\vspace{18\p@ plus18\p@}}
\def\submitted{\def\@submitted{\vspace{\baselineskip}%
     \noindent Submitted to: \journal\par}}
\def\@submitted{}
\def\nosections{\vspace{30\p@ plus12\p@ minus12\p@}
    \noindent\ignorespaces}
\def\ack{\ifletter\bigskip\noindent\ignorespaces\else
    \section*{Acknowledgments}\fi}

\newif\ifnumbysec
\def\theequation{\ifnumbysec
      \arabic{section}.\arabic{equation}\else
      \arabic{equation}\fi}
\def\eqnobysec{\numbysectrue\@addtoreset{equation}{section}}
\newenvironment{ceqno}{\begin{ceqno}}{\end{ceqno}}
\def\ceqno{\begin{equation}\begin{array}{@{}*{4}{l}}\dsty}
\def\endceqno{\end{array}\end{equation}}
\def\eqalign#1{\null\vcenter{\def\\{\cr}\openup\jot\m@th
  \ialign{\strut$\displaystyle{##}$\hfil&$\displaystyle{{}##}$\hfil
      \crcr#1\crcr}}\,}
\def\eqalignno#1{\displ@y \tabskip\z@skip
  \halign to\displaywidth{\hspace{5pc}$\@lign\displaystyle{##}$%
    \tabskip\z@skip
    &$\@lign\displaystyle{{}##}$\hfill\tabskip\@centering
    &\llap{$\@lign\hbox{\rm##}$}\tabskip\z@skip\crcr
    #1\crcr}}
\def\cases#1{%
     \left\{\,\vcenter{\def\\{\cr}\normalbaselines\openup1\jot\m@th%
     \ialign{\strut$\displaystyle{##}\hfil$&\tqs
     \rm##\hfil\crcr#1\crcr}}\right.}%
\def\tabular{\def\@halignto{}\@tabular}
\newcommand{\Table}[1]{\def\t@blecap{\caption{#1}}%
   \setcounter{tabtype}{1}\futurelet\next\t@bplace}
\newcommand{\widetable}[1]{\def\t@blecap{\caption{#1}}%
   \setcounter{tabtype}{2}\futurelet\next\t@bplace}
\newcommand{\fulltable}[1]{\def\t@blecap{\caption{#1}}%
   \setcounter{tabtype}{3}\futurelet\next\t@bplace}%
\def\t@bplace{\ifx\next[\let\next=\@tabpl
                 \else\let\next=\@tabnopl\fi\next}
\def\@tabpl[#1]{\begin{table}[#1]\@t@bsize}
\def\@tabnopl{\begin{table}\@t@bsize}
\def\@t@bsize{\ifnum\thetabtype=3\begin{varindent}{0pt}%
   \else\begin{varindent}{\mathindent}\fi
   \t@blecap\lineup\item[]
   \ifnum\thetabtype=1
        \begin{tabular}{@{}l*{15}{l}}
   \else\ifnum\thetabtype=2
        \begin{tabular*}{\indentedwidth}{@{}l*{15}{@{\extracolsep{0pt
plus12pt}}l}}
   \else\begin{tabular*}{\textwidth}{@{}l*{15}{@{\extracolsep{0pt plus12pt}}l}}
   \fi\fi}
\def\endtab{\ifnum\thetabtype=1\end{tabular}
   \else\end{tabular*}\fi\end{varindent}\end{table}}

\def\lineup{\def\0{\hbox{\phantom{\footnotesize\rm 0}}}%
    \def\m{\hbox{$\phantom{-}$}}%
    \def\-{\llap{$-$}}}
\long\def\@makecaption#1#2{\vskip 10\p@
 \ifnum\thefigtype=2\begin{varindent}{\@figindent}
 \item[]{\bf #1.} #2
 \end{varindent}\else
 \ifnum\thefigtype=3
 \footnotesize\rm{\bf #1.} #2\else
 \begin{indented}
 \item[]{\bf #1.} #2
 \end{indented}\fi\fi}
\newcommand{\Figure}[1]{\setcounter{figtype}{1}%
    \def\figspace{}\def\figcap{\caption{#1}}%
    \futurelet\next\@figplace}
\def\@figplace{\ifx\next[\let\next=\@figpl
                 \else\let\next=\@fignopl\fi\next}
\def\@figpl[#1]{\begin{figure}[#1]
   \figspace
   \figcap
   \end{figure}}
\def\@fignopl{\begin{figure}
   \figspace
   \figcap
   \end{figure}}

\newcommand{\sidecap}[3]{\setcounter{figtype}{2}%
    \setlength{\figdepth}{#1}\def\@figindent{#2}%
    \def\sidedc@p{\caption{#3}}%
    \futurelet\next\@sidecapplace}
\def\@sidecapplace{\ifx\next[\let\next=\@sidecappl
                 \else\let\next=\@sidecapnopl\fi\next}
\def\@sidecappl[#1]{\begin{figure}[#1]
    \vbox to\figdepth{\vfill
    \sidedc@p}%
    \setcounter{figtype}{1}\end{figure}}
\def\@sidecapnopl{\begin{figure}
    \vbox to\figdepth{\vfill
    \sidedc@p}%
    \setcounter{figtype}{1}\end{figure}}
\newcommand{\side}[3]{\setcounter{figtype}{3}%
    \setlength{\figdepth}{#1}\setlength{\figwidth}{15pc}
    \setlength{\secfigwidth}{15pc}
    \def\firstc@p{\caption{#2}}\def\secondc@p{\caption{#3}}
    \futurelet\next\@sideplace}
\def\@sideplace{\ifx\next[\let\next=\@sidepl
                 \else\let\next=\@sidenopl\fi\next}
\def\@sidepl[#1]{\begin{figure}[#1]
    \vspace*{1.5pc}\vspace*{\figdepth}
    \parbox[t]{\figwidth}{\firstc@p}\hspace*{1pc}%
    \parbox[t]{\secfigwidth}{\secondc@p}
    \setcounter{figtype}{1}\end{figure}}
\def\@sidenopl{\begin{figure}
    \vspace*{1.5pc}\vspace*{\figdepth}
    \parbox[t]{\figwidth}{\firstc@p}\hspace*{1pc}%
    \parbox[t]{\secfigwidth}{\secondc@p}
    \setcounter{figtype}{1}\end{figure}}
\newcommand{\varside}[4]{\setcounter{figtype}{3}%
    \setlength{\figdepth}{#1}\setlength{\figwidth}{#2}%
    \setlength{\secfigwidth}{30pc}
    \addtolength{\secfigwidth}{-\figwidth}
    \def\firstc@p{\caption{#3}}\def\secondc@p{\caption{#4}}
    \futurelet\next\@sideplace}

\newcounter{figure}
\def\thefigure{\@arabic\c@figure}
\def\fps@figure{htbp}
\def\ftype@figure{1}
\def\ext@figure{lof}
\def\fnum@figure{\figurename~\thefigure}
\def\figure{\@float{figure}}
\let\endfigure\end@float
\@namedef{figure*}{\@dblfloat{figure}}
\@namedef{endfigure*}{\end@dblfloat}
\newcounter{table}
\def\thetable{\@arabic\c@table}
\def\fps@table{htbp}
\def\ftype@table{2}
\def\ext@table{lot}
\def\fnum@table{\tablename~\thetable}
\def\table{\@float{table}}
\let\endtable\end@float
\@namedef{table*}{\@dblfloat{table}}
\@namedef{endtable*}{\end@dblfloat}

\def\thebibliography#1{\list
 {\hfil[\arabic{enumi}]}{\topsep=0\p@\parsep=0\p@
 \partopsep=0\p@\itemsep=0\p@
 \labelsep=5\p@\itemindent=-10\p@
 \settowidth\labelwidth{\footnotesize[#1]}%
 \leftmargin\labelwidth
 \advance\leftmargin\labelsep
 \advance\leftmargin -\itemindent
 \usecounter{enumi}}\footnotesize
 \def\newblock{\ }
 \sloppy\clubpenalty4000\widowpenalty4000
 \sfcode`\.=1000\relax}

\def\numrefs#1{\begin{thebibliography}{#1}}
\def\endnumrefs{\end{thebibliography}}

\def\thereferences{\list{}{\topsep=0\p@\parsep=0\p@
 \partopsep=0\p@\itemsep=0\p@\labelsep=0\p@\itemindent=-18\p@
\labelwidth=0\p@\leftmargin=18\p@
}\footnotesize\rm
\def\newblock{\ }
\sloppy\clubpenalty4000\widowpenalty4000
\sfcode`\.=1000\relax
}
\newenvironment{harvard}{\list{}{\topsep=0\p@\parsep=0\p@
\partopsep=0\p@\itemsep=0\p@\labelsep=0\p@\itemindent=-18\p@
\labelwidth=0\p@\leftmargin=18\p@
}\footnotesize\rm
\def\newblock{\ }
\sloppy\clubpenalty4000\widowpenalty4000
\sfcode`\.=1000\relax}{\endlist}
\def\refs{\begin{harvard}}
\def\endrefs{\end{harvard}}
\newenvironment{indented}{\begin{indented}}{\end{indented}}
\newenvironment{varindent}[1]{\begin{varindent}{#1}}{\end{varindent}}
\def\indented{\list{}{\itemsep=0\p@\labelsep=0\p@\itemindent=0\p@
   \labelwidth=0\p@\leftmargin=\mathindent\topsep=0\p@\partopsep=0\p@
   \parsep=0\p@\listparindent=15\p@}\footnotesize\rm}

\def\varindent#1{\setlength{\varind}{#1}%
   \list{}{\itemsep=0\p@\labelsep=0\p@\itemindent=0\p@
   \labelwidth=0\p@\leftmargin=\varind\topsep=0\p@\partopsep=0\p@
   \parsep=0\p@\listparindent=15\p@}\footnotesize\rm}

\def\tabnotes{\ifnum\thetabtype=1\end{tabular}\else\end{tabular*}\fi}
\def\endtabnotes{\end{varindent}\end{table}}

\newif\if@restonecol
\def\theindex{\@restonecoltrue\if@twocolumn\@restonecolfalse\fi
\columnseprule \z@
\columnsep 35\p@\twocolumn[\section*{\indexname}]%
    \@mkboth{{\indexname}}{{\indexname}}%
    \parindent\z@
    \parskip\z@ plus.3\p@\relax\let\item\@idxitem}
\def\@idxitem{\par\hangindent 30\p@}
\def\subitem{\par\hangindent 30\p@ \hspace*{10\p@}}
\def\subsubitem{\par\hangindent 30\p@ \hspace*{20\p@}}
\def\endtheindex{\if@restonecol\onecolumn\else\clearpage\fi}
\def\indexspace{\par \vskip 10\p@ plus 5\p@ minus 3\p@\relax}

\mark{{}{}}

\def\ps@headings{\let\@mkboth\markboth
 \def\@oddfoot{}%
 \def\@evenfoot{}%
 \def\@evenhead{\makebox[\mathindent][l]{\normalsize\rm \thepage}%
  \normalsize\it\rightmark\hfill}%
 \def\@oddhead{\makebox[\mathindent][r]{\hfill}{\normalsize\it\leftmark}\hfill
  \normalsize\rm\thepage}%
}%

\def\ps@myheadings{\let\@mkboth\markboth
 \def\@oddhead{\catchline}%
 \def\@oddfoot{\cpyrtline}%
 \def\@evenhead{}%
 \def\@evenfoot{}%
}

\def\today{\ifcase\month\or
 January\or February\or March\or April\or May\or June\or
 July\or August\or September\or October\or November\or December\fi
 \space\number\day, \number\year}

\def\@begintheorem#1#2{\rm \trivlist \item[\hskip \labelsep{\it #1\ #2.}]}
\def\@opargbegintheorem#1#2#3{\rm \trivlist
      \item[\hskip \labelsep{\it #1\ #2\ (#3).}]}

\def\p@LaTeX{{L\kern-.3em\lower.1em\hbox{$^{\rm A}$}\kern-.15em%
    T\kern-.1667em\lower.7ex\hbox{E}\kern-.125emX}}
\newcommand{\text}[1]{\mbox{#1}}

\newcommand{\nohyphens}{\hyphenpenalty=10000\exhyphenpenalty=10000}

\renewcommand{\d}{{\mathrm d}}
\renewcommand{\qquad}{\hspace*{25pt}}
\newcommand{\tqs}{\hspace*{25pt}}
\newcommand{\fl}{\hspace*{-\mathindent}}

\def\pt(#1){({\it #1\/})}
\newcommand{\dsty}{\displaystyle}
\newcommand{\tsty}{\textstyle}

\def\;{\protect\psemicolon}
\def\psemicolon{\relax\ifmmode\mskip\thickmuskip\else\kern .3333em\fi}

\newcommand{\opencirc}{\raisebox{2\p@}{\;\circle{5}}}

\newcommand{\fullcirc}{\raisebox{-2\p@}{\Large$\bullet$}}

\newcommand{\chain}{\mbox{--- $\cdot$ ---}}

\newcommand{\boldarrayrulewidth}{1\p@}

\def\bhline{\noalign{\ifnum0=`}\fi\hrule \@height
\boldarrayrulewidth \futurelet \@tempa\@xhline}

\def\@xhline{\ifx\@tempa\hline\vskip \doublerulesep\fi
      \ifnum0=`{\fi}}

\newcommand{\br}{\ms\bhline\ms}
\newcommand{\mr}{\ms\hline\ms}
\newcommand{\ms}{\noalign{\vspace{3\p@ plus2\p@ minus1\p@}}}
\newcommand{\bs}{\noalign{\vspace{6\p@ plus2\p@ minus2\p@}}}
\newcommand{\ns}{\noalign{\vspace{-3\p@ plus-1\p@ minus-1\p@}}}
\newcommand{\es}{\noalign{\vspace{6\p@ plus2\p@ minus2\p@}}\displaystyle}
\newcommand{\etal}{{\it et al\/}\ }

\newcommand{\JPA}{{\em J. Phys. A: Math. Gen.} }

\newcommand{\PRL}{{\em Phys. Rev. Lett.} }

\ps@headings  \onecolumn
\makeatother
\begin{document}
\jl{1}
\input{epsf.sty}
\begin{flushright}
SISSA Ref. 39/97/EP
\end{flushright}
\vspace*{0.5cm}
\begin{center}
   {\bf \Large 
First-order phase transitions in one-dimensional steady states
 	}
   \\[15mm]
Peter F. Arndt$\mbox{}^\star$,
Thomas Heinzel$\mbox{}^\star$
and Vladimir Rittenberg$\mbox{}^\diamond$
\\[7mm]
{SISSA, Via Beirut 2--4, 34014 Trieste, Italy\\[3mm]
permanent address:
Physikalisches Institut, Nu{\ss}allee 12,
53115 Bonn, Germany}
\\[1.5cm]
\end{center}
\renewcommand{\thefootnote}{\arabic{footnote}}
\addtocounter{footnote}{-1}
\vspace*{2mm}
%
The steady states of the two-species (positive and negative particles)
asymmetric exclusion model of Evans, Foster, Godr{\`e}che and Mukamel are
studied  using Monte Carlo simulations. 
We show that mean-field theory does not
give the correct phase diagram. 
On the first-order phase transition line which separates
the $CP$-symmetric phase from the broken phase, the density profiles can be
understood through an unexpected pattern of shocks. 
In the broken phase the
free energy functional is not a convex function but looks like a standard
Ginzburg-Landau picture. 
If a symmetry breaking term is introduced in the
boundaries the Ginzburg-Landau picture remains and one obtains spinodal points.
The spectrum of the hamiltonian associated with the master equation was
studied using numerical diagonalization. 
There are massless excitations on the
first-order phase transition line with a dynamical critical exponent $z=2$
as expected from the existence of shocks and at the spinodal points 
where we find $z=1$.
It is for the first time that this value which characterizes conformal
invariant equilibrium problems appears in stochastic processes.
\vspace*{1.5cm}
\begin{flushleft}
cond-mat/9706114
\\
June 1997\\[1cm]
$\mbox{}^\star$ work supported by the DAAD programme HSP II--AUFE\\
$\mbox{}^\diamond$  work done with partial support of the EC TMR programme,
grant FMRX-CT96-0012
\end{flushleft}
\thispagestyle{empty}
\mbox{}
\newpage
\setcounter{page}{1}
\eqnobysec
\def\bc{\beta_{\rm crit}}
\def\hc{h_{\rm crit}}
\def\Ts{T_{\rm short}}
\def\Tl{T_{\rm long}}
\def\hp{\frac{1+h}{2}}
\def\hm{\frac{1-h}{2}}
\def\d{{\rm d\/}}
\def\np{\newpage}
\def\qua{{
        \setlength{\unitlength}{0.1mm}
        \begin{picture}(24,20)(-2,1)
        \put(0,0){\line(1,0){20}}
        \put(0,0){\line(0,1){20}}
        \put(20,20){\line(-1,0){20}}
        \put(20,20){\line(0,-1){20}}
        \end{picture}
        }       }
\def\pb{p}
\def\mb{m}
\def\hb{n}
\def\de{d}
\def\hc{h_{\rm crit}}
\def\c{c}
\def\erf{\;\mbox{erf}}
\def\FEF{FEF}
\section{Introduction}
Several years ago Krug \cite{ci} suggested the existence of boundary induced
phase transitions in one-dimensional steady states. 
A line of first-order
phase transitions (the so-called coexistence line) was found in the one-species
asymmetric exclusion model \cite{cii,ciii,civ} 
in which particles (call them positive)
hop among vacancies (call them negative particles). 
On the coexistence
line, the system is $CP$-symmetric ($C$ corresponds to changing the sign of
the particles, $P$ is the parity operation) and the symmetry is
spontaneously broken (some aspects of this phase transition which are
relevant to this paper are reviewed in Appendix A). 
Since we want to compare
the nature of the phase transitions in equilibrium and non-equilibrium
phenomena, the phase transition just described would correspond to the
two-dimensional $n$-state Potts model \cite{cv} 
with $n>4$ at the critical
temperature. 
The exclusion model does not have the equivalent of the low
temperature domain which corresponds to the broken phase.

The two-species model of Evans, Foster, Godr{\`e}che 
and Mukamel \cite{cvi,cvii} 
presents a broken phase 
(the equivalent of the low temperature domain of the $n$-states
Potts models) and this motivated us to study this model in detail. In our
investigation we got a few surprises.

We will just reproduce from Ref. \cite{cvii} 
the definition of the model,
introducing also symmetry breaking boundary terms, and send
the reader to the same paper in order to find out why the model is
physically relevant.
Each site of a one-dimensional lattice with $L$ sites may be occupied by a
positive particle or a negative particle or be empty. 
In each
infinitesimal time step $\d t$ the following events may occur at each
nearest-neighbor sites $k$, $k+1$:
\begin{eqnarray}
\label{einti}
(+)_k\,(\,0\,)_{k+1}\,\rightarrow (\,0\,)_k\,(+)_{k+1}
\nonumber\\
(+)_k\,(-)_{k+1}\,\rightarrow (-)_k\,(+)_{k+1}
\\
(\,0\,)_k\,(-)_{k+1}\,\rightarrow (-)_k\,(\,0\,)_{k+1}
\nonumber
\end{eqnarray}
all with probability $\d t$. 
Where $(+)_k$, $(-)_k$ and $(\,0\,)_k$ indicate a positive
particle, a negative particle or a vacancy at the site $k$.
In the same time step ${\d t}$ the following events may occur at the left
boundary ($k=1$)
\begin{eqnarray}
\label{eintii}
(\,0\,)_1 \rightarrow (+)_1 
\quad\mbox{with probability }\alpha\; \d t
\nonumber\\
(-)_1 \rightarrow (\,0\,)_1 
\quad\mbox{with probability }\beta(1-h)\; \d t
\end{eqnarray}
and at the right boundary ($k=L$)
\begin{eqnarray}
\label{eintiii}
(\,0\,)_L \rightarrow (-)_L 
\quad\mbox{with probability }\alpha \;\d t
\nonumber\\
(+)_L \rightarrow (\,0\,)_L 
\quad\mbox{with probability }\beta(1+h) \;\d t\;  .
\end{eqnarray}
One notices that for $h=0$ the probabilities are $CP$-invariant, the case
$h$ bigger than zero will be considered in Section 4.

The time evolution of the system is given by a master equation or its
equivalent imaginary time Schr{\"o}dinger equation \cite{cix}
\begin{equation}
\label{eintiv}
\frac{\d}{\d t} |P>=-H\,|P>
\end{equation}
where $|P>$ is the probability vector and $H$ is the hamiltonian of a
one-dimensional quantum chain which is given in Appendix B. 
For most of
this paper we will be interested in the properties of the steady state.

Let us introduce some notations: $p(k)$, $m(k)$ and $v(k)$ 
will denote the steady state density
of positive, negative particles and vacancies at the site $k$. 
Their average
values being $p$, $m$ respectively $v$. 
It is also useful to define the
quantities
\begin{eqnarray}
\label{eintv}
d(k)=p(k)-m(k) 
\nonumber\\
s(k)=(p(k)+m(k))/2\, ,
\end{eqnarray}
their average values being $d$ and $s$.

%
%
\begin{figure}
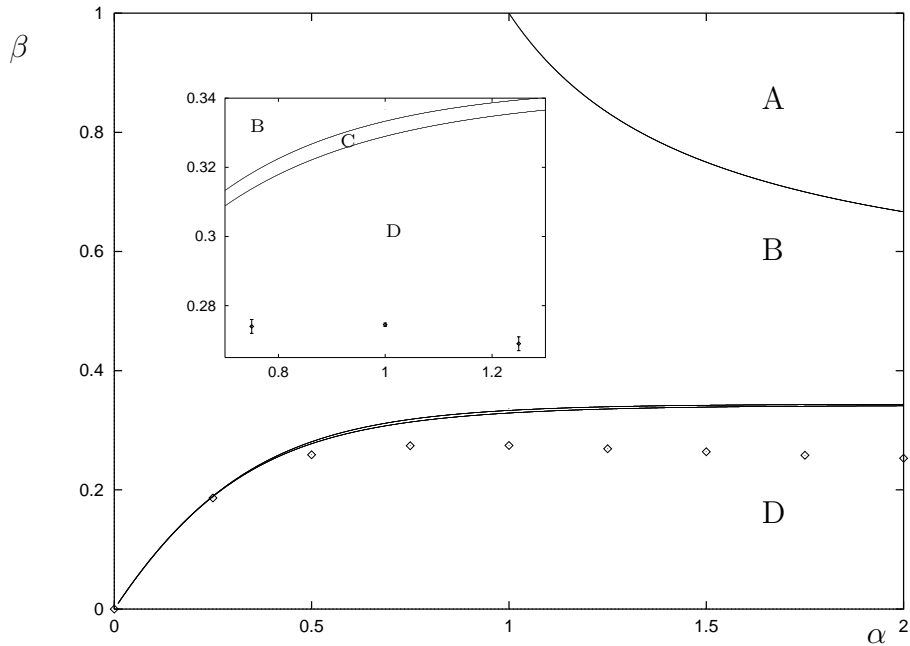

\def\sgr{\scriptstyle}
\setlength{\unitlength}{1mm}
\def\setl{\setlength\epsfxsize{12cm}}
\begin{picture}(155,90)(-20,-2)
\put(114,-1){\makebox{$\alpha$}}
\put(0,77){\makebox{$\beta$}}
\put(100,70){\makebox{A}}
\put(100,50){\makebox{B}}
\put(32,67){\makebox{\tiny B}}
\put(44,65){\makebox{\tiny C}}
\put(50,53){\makebox{\tiny D}}
\put(100,15){\makebox{D}}
\put(0,0){
        \makebox{
                \setl
                \epsfbox{f1.epsf}}
        }
\put(15,34){
        \makebox{
               \setlength\epsfxsize{5.5cm}
               \epsfbox{f2.epsf}}
       }
\end{picture}
\caption{
\label{figvii}
Phase diagram for h=0.
The phases A, B, C and D are obtained in mean-field and defined in the text.
Since the C phase is very narrow it is shown in an insert.
The diamonds describe the boundary between the phase $\tilde{\rm B}$ 
and the phase
$\tilde{\rm D}$ 
as obtained in Monte Carlo simulations. (No phase C is observed.)
         }
\end{figure}

The phase diagram corresponding to the steady state was obtained in
mean-field theory 
for the $CP$-symmetric case \cite{cvii} and is shown in Fig.\ref{figvii}. 
One distinguishes four phases, we
will give the average densities of particles in each one.
The power law phase (A):
\begin{equation}
\label{eintvi}
p=m=\frac12, \qquad v=0
\end{equation}
The low densities ($p=m<1/2$) symmetric phase (B):
\begin{equation}
\label{eintvii}
p=m=\frac{\alpha\beta}{\alpha+\beta}<\frac12
\end{equation}
The low densities ($p,m <1/2$) broken phase (C): 
Here mean-field gives three
solutions,
A symmetric unstable solution
\begin{equation}
\label{eintviii}
p=m=\frac{\alpha\beta}{\alpha+\beta}<\frac12
\quad \mbox{(unstable)},
\end{equation}
and two stable solutions
\begin{eqnarray}
\label{eintix}
s=2\left( 1-\frac{\alpha\beta}{\alpha-\beta}\right)
\nonumber\\
d=\pm 2 \left[ (s-1)(\frac{\alpha\beta}{\alpha-\beta}-s)\right]^{\frac12} \, .
\end{eqnarray}
The low density/high density phase (D):
mean-field gives again three
solutions. A symmetric unstable one which is again described by equation
(\ref{eintviii}) like in phase C, 
a second solution in which the positive particles have a high
density ($p>1/2$) and the negative particles have a low density ($m<1/2$):
\begin{eqnarray}
\label{eintx}
p=1-\beta
\nonumber\\
m=\frac{1+\alpha}{2}-\frac12 \left[ (1+\alpha)^2 -4\alpha\beta \right]^{\frac12}
\end{eqnarray}
and a third solution in which in Eq.(\ref{eintx}) $p$ is exchanged with $m$.
For reasons which will become apparent immediately we have checked which
solutions are stable with respect to the mean-field 
dynamics.

In mean-field theory the transitions between the phases A and B respectively B
and C are continuous and the transition between C and D is first-order. 

An exact calculation
\cite{cvii} using the matrix product ansatz for the 
wave function on the line $\beta=1$
has confirmed the existence of phases A and B. Limited
Monte Carlo simulations \cite{cvii} 
showed a strong similarity between the true and mean-field
phase diagrams for $\alpha=1$.
Since for the two-species model 
\cite{cii,ciii,civ,cviii,cix}
the phase diagram obtained by mean-field
is exact, it might be tempting to conclude that in the present
case the mean-field phase diagram is exact.

In Section 2 we show a detailed Monte Carlo analysis of the model with an
unexpected result. The D 
phase exists only up to a line going through the
diamonds in Fig.\ref{figvii} 
where a first-order phase transition takes place. 
We will denote the domain under this line by $\tilde{\rm D}$. 
The C phase does not exist,
the B phase extends down to the $\tilde{\rm D}$ phase and will 
be denoted by $\tilde{\rm B}$ (see Fig.\ref{figxxx}). 
In this picture the role of
mean-field is quite perverse. 
In the $\tilde{\rm D}$ domain (broken phase) one has the
two solutions described by equation (\ref{eintx}).
Above the line of diamonds the system picks up the symmetric unstable
solution given by equation (\ref{eintviii}). 
This solution coincides with the one
which defines phase B.

A useful tool in this analysis is played by the free energy functional
which is defined as follows \cite{cx}:
Take an order parameter $w$ and let $P(L,w)$ be the probability to have $w$ 
for a lattice with $L$ sites. 
Defining
\begin{equation}
\label{eintxi}
f_L(w)=-\frac1{L}\log P(L,w)
\end{equation}
the free energy functional (\FEF) is
\begin{equation}
\label{eintxii}
f(w)=\lim_{L\rightarrow\infty} f_L(w)\; .
\end{equation}
For equilibrium systems this is a convex function. 
As we are
going to see this is not true for steady states 
which do not describe Gibbs ensembles.
In the $\tilde{\rm D}$ domain if we
take $d$ as order parameter the \FEF\  looks
like a Ginzburg-Landau picture in the case of spontaneous breaking
of a symmetry \cite{GiLa}. 
On the first-order transition line separating the
phases $\tilde{\rm B}$ and $\tilde{\rm D}$ the \FEF\ is convex, 
looking like the \FEF\ 
corresponding to a
first-order phase transition in equilibrium phenomena.
Above the diamond-line the \FEF\ is a convex function
with one minimum only.
We have also
looked at the spectrum of the hamiltonian (see Appendix B) on
the separation line. 
It is gapless with a critical dynamical exponent \cite{cxi}
$z=2$. This suggests, from the experience with the one-species model, the
existence of shocks \cite{cxii,cxiii}.
The problems of convergence coming from 
the definition (\ref{eintxii}) are discussed
in  Appendix A.
(Some of the ideas
developed here are already in the paper of Bennett and Grinstein \cite{cxiv}). 

In Section 3 we study in detail (for $\alpha =1$)
the nature of the first-order 
phase transition. 
A novel feature comes from the fact that we have not one
but two order parameters ($p$ and $m$ or $s$ and $d$). 
In the $p$-$m$-plane the minima 
of the \FEF\ lie on a boomerang like figure. 
We show how this figure can be
obtained by an interesting combination of shocks. 
The existence of shocks
allows the prediction of different density profiles which are indeed
observed in several Monte Carlo simulations. 

In Section 4 we consider the effect of a symmetry breaking term on the process
(we take $h$ bigger than zero in Eqs.(\ref{eintii}--\ref{eintiii})).
We show that the \FEF s look like
text book Ginzburg-Landau pictures including the
existence of spinodal points in the $\tilde{\rm D}$ phase. 
As is well known \cite{cxvi} such points do not
exist in equilibrium phenomena. 
The mean-field calculations are presented elsewhere
\cite{cxxix}. They are compared 
here with the Monte Carlo simulations. It looks like mean-field
theory gives a good approximation to the data only for small
values of $\beta$ and large values of $h$.

Moreover, we show that the spectrum
of the hamiltonian defined in Appendix B shows massless exitations at the
spinodal point with a dynamical critical exponent $z=1$. 
This is a very
surprising result for stochastic processes. This exponent is typical for
continuous equilibrium phase transitions.
In Section 5 we present our conclusions.
%
%
%
\def\figii{
\begin{figure}[b]
\def\sgr{\scriptstyle}
\setlength{\unitlength}{1mm}
\def\setl{ \setlength\epsfxsize{12cm}}
\begin{picture}(155,90)(-20,-2)
\put(114,-1){\makebox{$\de$}}
\put(0,77){\makebox{$f_L(\de)$}}
\put(70,10){\makebox{$\beta=0.264$}}
\put(70,50){\makebox{$\beta=0.290$}}
\put(0,77){\makebox{$$}}
\put(0,0){
        \makebox{
                \setl
                \epsfbox{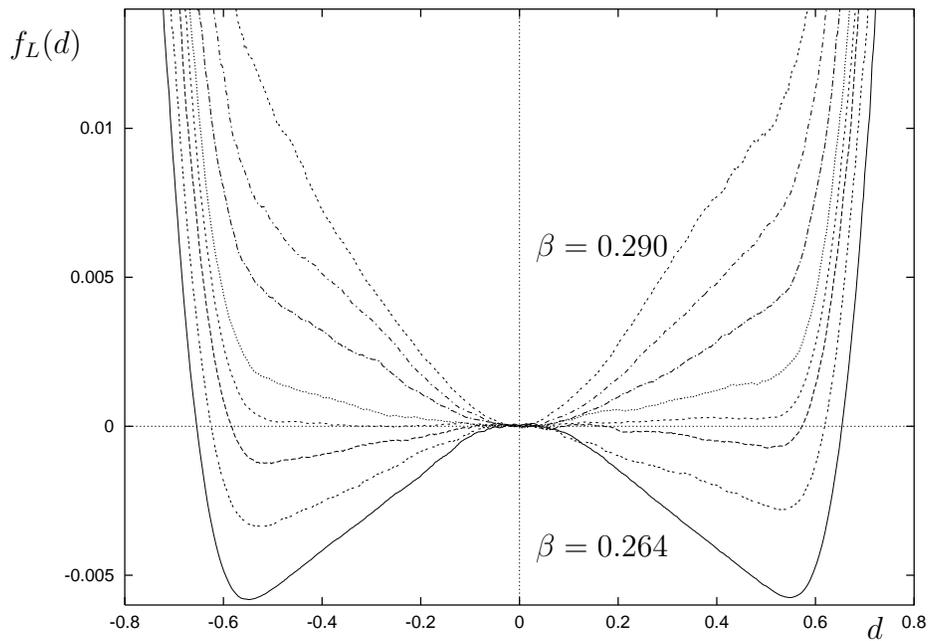}}
        }
\end{picture}
\caption{
\label{figiv}
$f_L(\de)$ 
for $\beta=$0.264, 0.268, 0.272, 0.274, 0.276, 0.280, 0.285, 0.290
and $L=400$ sites.
The values of $f_L(\de)$ are shifted by their value $f_L(0)$.
         }
\end{figure}
}
%
%
\def\tabi{
\begin{table}
\caption{The critical points for various $\alpha$.}
\begin{indented}
\item[]
\begin{tabular}{@{}llllllllllllllllllllllll}
\br
$\alpha$&$\bc$\qquad\qquad&$\alpha$&$\bc$
\\
\mr
0.125	&0.11(2)        &1.00   &0.275(1)\\
0.25	&0.186(2)	&1.25	&0.269(2)\\
0.375	&0.234(2)	&1.50   &0.264(2)\\
0.50	&0.259(2)	&1.75   &0.258(2)\\
0.75	&0.274(2)	&2.00	&0.253(2)\\
\br
\end{tabular}
\end{indented}
\end{table}
}
%
%
\def\figiii{
\begin{figure}
\def\sgr{\scriptstyle}
\setlength{\unitlength}{1mm}
\def\setl{ \setlength\epsfxsize{12cm}}
\begin{picture}(155,90)(-20,-2)
\put(114,-1){\makebox{$\de$}}
\put(0,77){\makebox{$f_L(\de)$}}
\put(0,0){
        \makebox{
                \setl
                \epsfbox{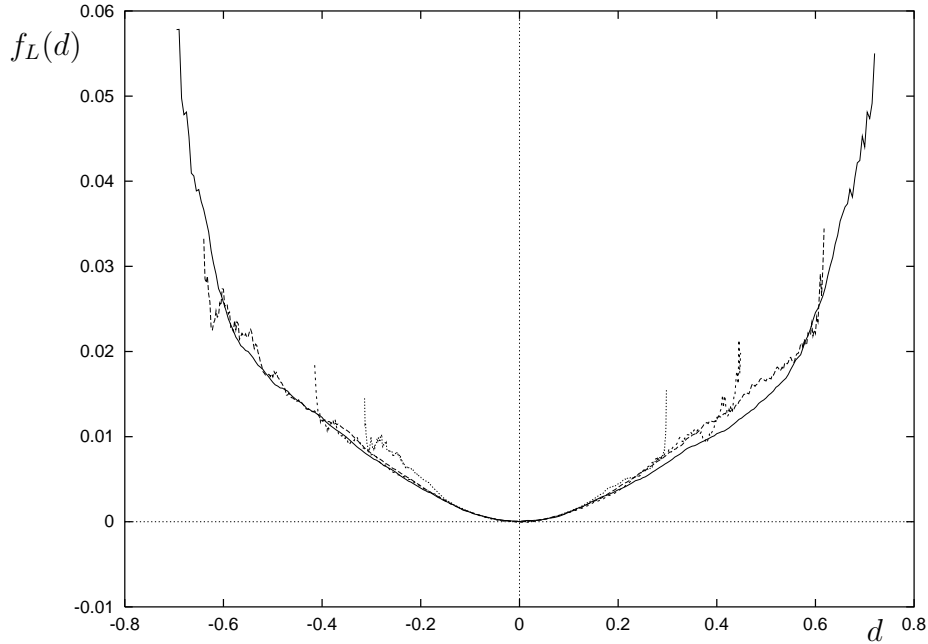}}
        }
\end{picture}
\caption{
\label{figvi}
Free energy functional for $\beta=0.3$: 
$f_L(\de)$ for
$L=$200, 400, 600 and 1000 sites.
The values of $f_L(\de)$ for different $L$ are shifted by $f_L(0)$.
The little vertical stretches of the curves are due to the
limited CPU time of the Monte Carlo runs.
         }
\end{figure}
}
%
%
\def\figiv{
\begin{figure}[tb]
\def\sgr{\scriptstyle}
\setlength{\unitlength}{1mm}
\def\setl{ \setlength\epsfxsize{12cm}}
\begin{picture}(155,90)(-20,-2)
\put(114,-1){\makebox{$\beta$}}
\put(0,77){\makebox{$|\de|$}}
\put(0,0){
        \makebox{
                \setl
                \epsfbox{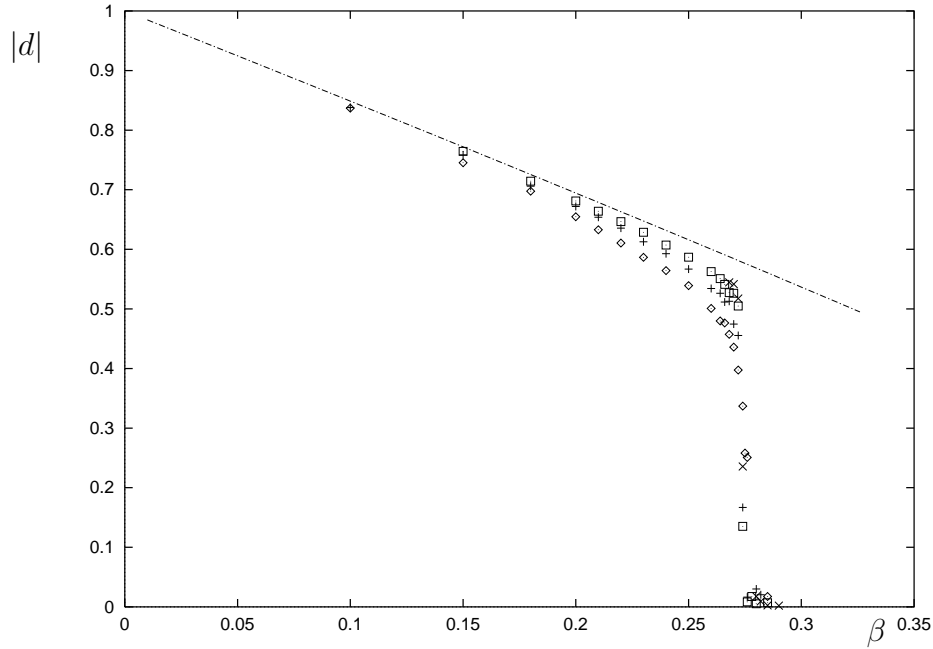}}
        }
\end{picture}
\caption{
\label{figv}
Minima of $f_L(\de)$ for
$L=100(\diamond), 200(+), 400(\protect\qua), 600(\times)$ sites.
The curve (\chain) gives the mean-field prediction
obtained from Eq.(\protect\ref{eintx}).
         }
\end{figure}
}
%
\def\figv{
\begin{figure}[tb]
\def\sgr{\scriptstyle}
\setlength{\unitlength}{1mm}
\def\setl{ \setlength\epsfxsize{12cm}}
\begin{picture}(155,90)(-20,-2)
\put(114,-1){\makebox{$\beta$}}
\put(0,77){\makebox{$s$}}
\put(0,0){
        \makebox{
                \setl
                \epsfbox{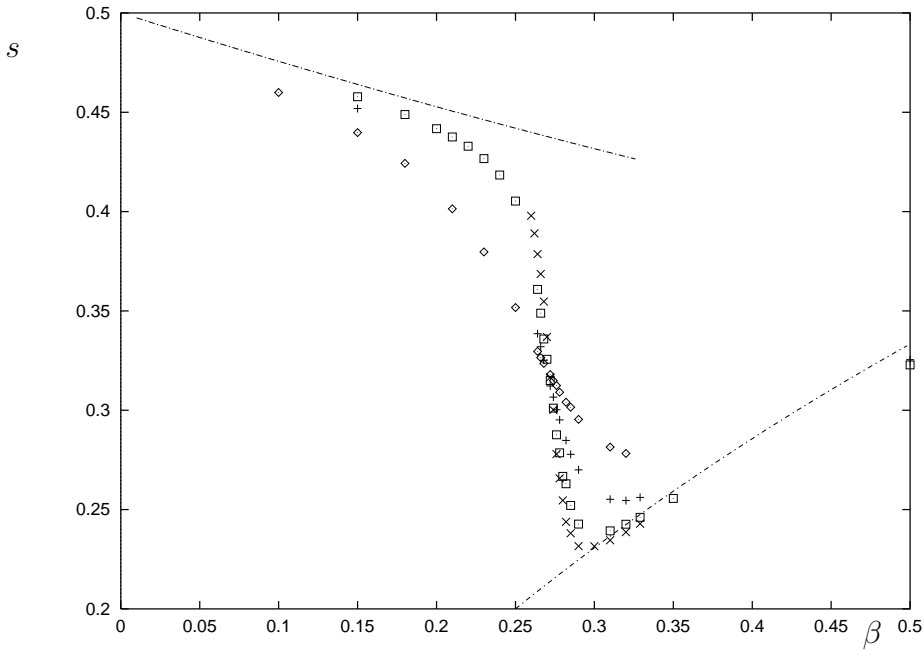}}
        }
\end{picture}
\caption{
\label{figiii}
Minima of $f_L(s)$ ($s=(p+m)/2$) as a function of $\beta$
for $L=100(\diamond),200(+),400(\protect\qua),600(\times)$ sites.
The upper curve corresponds to Eq.(\protect\ref{essi}) (broken phase),
the lower one to Eq.(\protect\ref{essii}) (unbroken phase).
         }
\end{figure}
}
%
%
%
%
\def\figvi{
\begin{figure}[tb]
\def\sgr{\scriptstyle}
\setlength{\unitlength}{1mm}
\def\setl{ \setlength\epsfxsize{12cm}}
\begin{picture}(155,90)(-20,-2)
\put(114,-1){\makebox{$\beta$}}
\put(0,77){\makebox{$E_{\infty}$}}
\put(0,0){
        \makebox{
                \setl
                \epsfbox{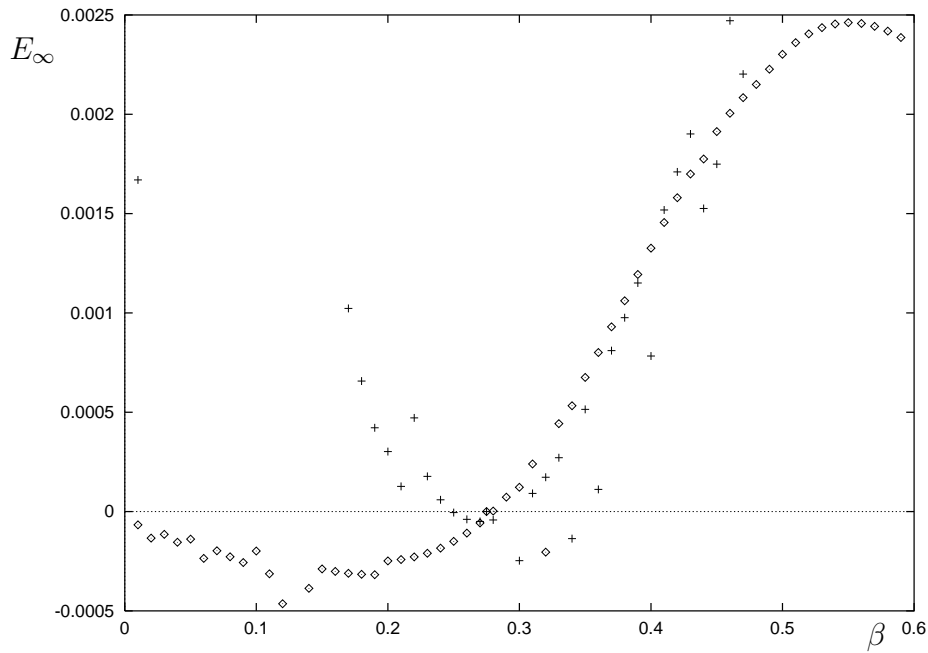}}
        }
\end{picture}
\caption{
\label{figi}
Estimates of the first ($\diamond$) and second (+) excitations
of the $CP$-symmetric hamiltonian (Eq. (\protect\ref{ham}) with $h=0$).
The estimates were obtained from the spectra of the 
hamiltonian up to $L=11$ by extrapolating the 
results in standard ways.
Since all eigenvalues of $H$ are positive their errors
are estimated to be 0.0003.
         }
\end{figure}
}
%
\section{The phase transition between the broken and unbroken phases}
\figii
\figiii
\figiv
\figv
\figvi
In order to clarify the phase structure of the model, let us take
$\alpha=1$.
(This is a vertical line in Fig.\ref{figvii}). 
If one wants to compare
equilibrium with non-equilibrium phenomena, it is useful to have in mind
the two-dimensional $n$-state Potts model with $n>4$ and interpret 
the parameter $\beta$ as a
temperature. Mean-field gives the following values of $\beta$ for the
separations among the various phases:
\begin{itemize}
\item
$\beta =0.3289$ (between phase D and phase C)
\item
$\beta= 0.3333 = 1/3$ (between phase C and phase B) 
\item
$\beta= 1$ (between phase B and phase A)
\end{itemize}
In order to check the mean-field predictions we have done Monte
Carlo simulations to get $f_L(\de)$
taking $\de$ (the difference between the averages of positive and
negative particles) as order parameter. 

In Fig.\ref{figiv} we show our results for
$L=400$ and different values of $\beta$ (all belonging to the D phase). 
In order
to present the data in a clearer way, we have substracted the values of $f_L(0)$
from $f_L(\de)$. The data shown in Fig.\ref{figiv} are interesting for two reasons:
the
shape of the \FEF\ and the failure of mean-field predictions.
First we notice that a dramatic change occurs around $\beta=0.274$.
As
will be shown below the precise value is $\bc=0.275$. Below $\bc$ the
\FEF\ is not a convex function, it has the behavior of a text-book 
Ginzburg-Landau picture, and shows two minima at the values given by
mean-field (see Eq.(\ref{eintx}) and the definition (\ref{eintv})).

At $\bc$ the \FEF\ behaves like the \FEF\ of an equilibrium first-order phase
transition. For $\beta>\bc$ the \FEF\ has the 
expected behavior for a disordered
phase. Notice that all the changes in the behavior of $f_L(\de)$ occur at
values of $\beta$ below the value 0.3289 where, according to mean-field, the
transition between the C and D phases is supposed to take place.

In order to make sure that we are not dealing with finite-size effects we have
computed $f_L(\de)$ for $\beta=0.3$ (still below 0.3289) for various number of
sites up to $L=1000$. The data are shown in Fig.\ref{figvi} 
(we have substracted the
value $f_L(0)$). One can see that $f_L(\de)$ is practically independent of $L$
(the $L$ dependence is in $f_L(0)$ -- see Appendix A). 
This implies that one does
not have finite-size effects, and that $f_L(\de)-f_L(0)$ 
is a convex function
with one minimum as one expects 
to have if one is in the ``disordered'' phase.

We now present more data which show what goes wrong
with the mean-field prediction. In Fig.\ref{figv} 
we show the positions of the minima
of $f_L(\de)$ for various numbers of sites and various values of $\beta$. 
One can
observe that below $\bc$ the minima converge to their mean-field values
obtained from Eq.(\ref{eintx}) and that a 
dramatic change takes place at $\bc$. 
Above
$\bc$ the $CP$ symmetry is not broken anymore ($\de=0$). 

In Fig.\ref{figiii} we show the minima of $f_L(s)$ as a function of $\beta$ 
for various $L$. 
The upper curve is given by the average value of $s$ for
the two phases (low/high density).
This value is obtained from Eq.(\ref{eintx}):
\begin{equation}
\label{essi}
s=\frac{1-\beta}{2}+\frac{1+\alpha}{4}-\frac14 
{\left[ (1+\alpha)^2-4\alpha\beta \right]}^{\frac12}
\end{equation}
The lower curve corresponds to the symmetric solution (\ref{eintviii}),
unstable below $\beta=1/3$:
\begin{equation}
\label{essii}
s=\frac{\alpha\beta}{\alpha+\beta}
\end{equation}
As one can see from Fig.\ref{figiii}, 
below $\bc$ the data converge to the values of $s$
given by Eq.(\ref{essi})  
and above $\bc$ to the values of $s$ given by Eq.(\ref{essii}).
This is very interesting because it means that fluctuations turn the unstable
solution into a stable one and one obtains a first-order phase transition. 
The
best estimate for $\bc$ comes from the ``fixed point'' seen in Fig.\ref{figiii}, 
one
gets $\bc=0.275\pm 0.001$ for $\alpha=1$. 
By the same method but not with the same patience we 
have computed
the values of $\bc$
for other values of $\alpha$. 
The values are given in Table 1.
\tabi

In the correct phase diagram which replaces Fig.1 one has only three phases:
the phase A like in Fig.\ref{figvii}, the phase $\tilde{\mbox{B}}$
(the unbroken phase) which extends 
down to the diamond-line in Fig.\ref{figvii} 
and finally the phase
$\tilde{\mbox{D}}$ under the diamond-line (see Fig.\ref{figxxx}).
We have not checked if the seperation between the A and 
$\tilde{\mbox{B}}$ phases is properly
given by mean-field but we believe this to be the case 
because of the analytical calculation
on the $\beta=1$ line done in \cite{cvii}. 

A final test of our picture can be obtained looking at the spectrum of the
hamiltonian (explicitly given in Appendix B). 
We consider the first three levels and
take $\alpha=1$. Since the ground state has energy zero, 
the values of the energies of the next two levels, 
denoted by $E_1$ and $E_2$, 
coincide with the energy gaps.
If our picture is correct $E_1$ should
vanish for large values of $L$ for 
$\beta<\bc$ since in the large $L$ limit
one needs two states, the ground state and the first excited state, to
describe the two vacua. $E_2$ should be finite. For $\beta>\bc$
both $E_1$ and $E_2$ should be finite.
Using the 
modified Arnoldi algorithm of 
Ref.\cite{uli} we have calculated, for various
values of $\beta$ and chain lengths (up to $L=11$) the values of 
$E_1$ and $E_2$.
In order to get their large $L$ limit we have used 
Bulirsch-Stoer approximants  
\cite{BST}. The results are shown in Fig.\ref{figi} 
which confirms the existence of a
critical point around $\beta=0.275$.

If one takes $\beta=\bc$ one can estimate the dynamical critical
exponent $z$ defined as follows:
\begin{equation}
\label{essiii}
\lim_{L\rightarrow\infty} L^z E_1(L)=\mbox{const.}>0
\end{equation}
where $E_1(L)$ represents the energy of the first exited state for a chain
of length $L$. We find $z=2.0\pm0.1$. This result is also 
interesting because from
our experience with the one-species problem 
(see Appendix A) a value of $z=2$
suggests the existence of shocks. In the next section we are going to find
them.
\def\figviii{
\begin{figure}
\def\sgr{\scriptstyle}
\setlength{\unitlength}{1mm}
\def\setl{\setlength\epsfxsize{12cm}}
\def\kreuz{\begin{picture}(0,0)
		\thicklines
		\put(-2,-2){\line(1,1){4}}
		\put(-2,2){\line(1,-1){4}}
		\thinlines
		\end{picture}
	}
\begin{picture}(155,130)(-20,-2)
\put(114,-1){\makebox{$\pb$}}
\put(0,114){\makebox{$\mb$}}
\put(30,109){\kreuz}
\put(20,108){\makebox{R}}
\put(30,47){\kreuz}
\put(20,45.5){\makebox{S}}
\put(39,38){\kreuz}
\put(34,41){\makebox{T}}
\put(48,29){\kreuz}
\put(46,19){\makebox{U}}
\put(110,29){\kreuz}
\put(108.5,19){\makebox{V}}
\put(0,-20){
        \makebox{
                \setl
                \epsfbox{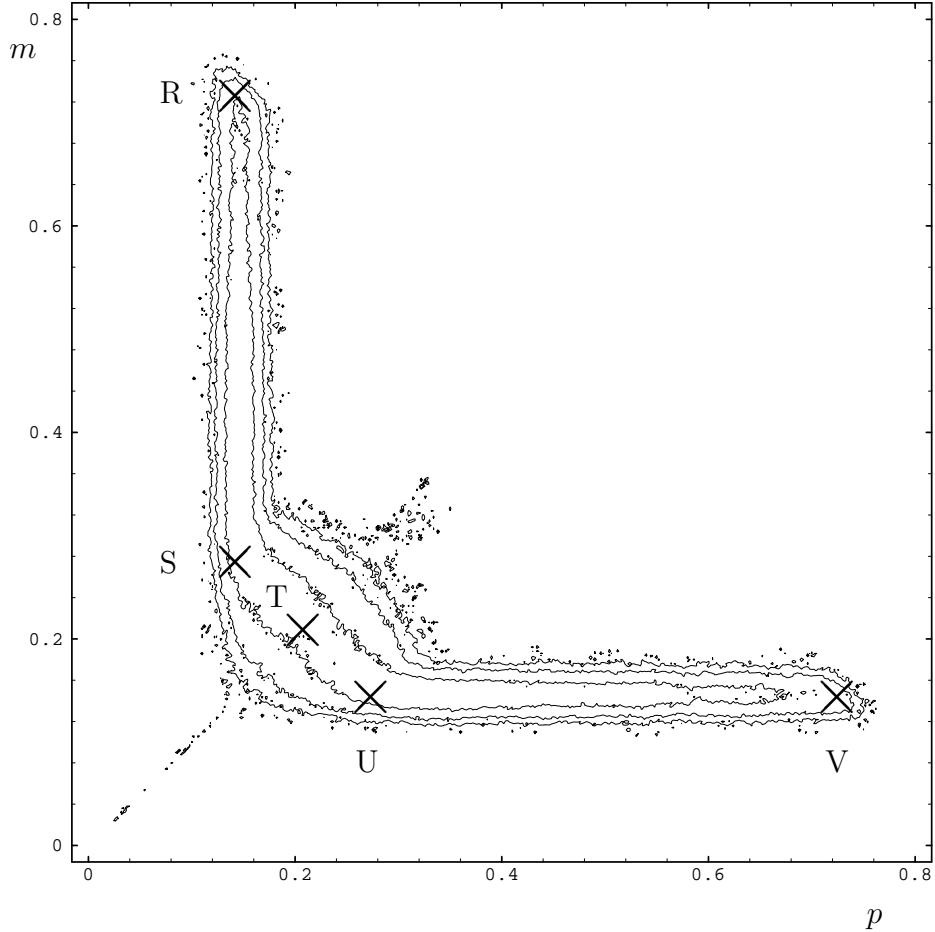}}
        }
\end{picture}
\caption{
\label{figviii}
Contour lines of  $f_L(\pb,\mb)$ for 
$\beta=0.275$ and $L=1000$ sites.
The ``height'' difference between lines is constant.
         }
\end{figure}
}
%
\def\figx{
\begin{figure}[b]
\def\sgr{\scriptstyle}
\setlength{\unitlength}{1mm}
\def\setl{ \setlength\epsfxsize{12cm}}
\begin{picture}(155,90)(-20,-2)
\put(114,-1){\makebox{$z$}}
\put(0,77){\makebox{$m(z)$}}
\put(0,0){
        \makebox{
                \setl
                \epsfbox{f9.epsf}}
        }
\end{picture}
\caption{
\label{figx}
Density profile of the negative particles for $L=50(\protect\qua),
100(\diamond), 200(+)$ at the phase transition $\beta=0.275$.
The straight line is given by Eq.(\protect\ref{esssiii}).
         }
\end{figure}
}
%
%
\def\figxi{
\begin{figure}
\def\sgr{\scriptstyle}
\setlength{\unitlength}{1mm}
\def\setl{ \setlength\epsfxsize{12cm}}
\begin{picture}(155,90)(-20,-2)
\put(114,-1){\makebox{$z$}}
\put(0,77){\makebox{$m(z)$}}
\put(0,0){
        \makebox{
                \setl
                \epsfbox{f10.epsf}}
        }
\end{picture}
\caption{
\label{figxi}
Density profile of the negative particles for $L=50(\diamond),
100(+), 200(\protect\qua)$ at $\beta=0.26$, below the phase transition.
The line gives the mean-field
prediction, i.e. the average of the densities of Eq.(\protect\ref{eintx}).
         }
\end{figure}
}
%
\def\figxii{
\begin{figure}
\def\sgr{\scriptstyle}
\setlength{\unitlength}{1mm}
\def\setl{ \setlength\epsfxsize{12cm}}
\begin{picture}(155,90)(-20,-2)
\put(114,-1){\makebox{$z$}}
\put(0,77){\makebox{$m(z)$}}
\put(0,0){
        \makebox{
                \setl
                \epsfbox{f11.epsf}}
        }
\end{picture}
\caption{
\label{figxii}
Density profile of the negative particles for $L=50(\diamond),
100(+), 200(\protect\qua)$ at $\beta=0.30$, above the phase transition. 
The line gives the (unstable) mean-field solution (\protect\ref{eintviii}).
         }
\end{figure}
}
%
%
\def\figxv{
\begin{figure}
\def\sgr{\scriptstyle}
\setlength{\unitlength}{1mm}
\def\setl{ \setlength\epsfxsize{12cm}}
\begin{picture}(155,90)(-20,-2)
\put(114,-3){\makebox{$y=\frac{k-k_0}{\sqrt{L}}$}}
\put(55,40){\makebox{$p(y)$}}
\put(55,11){\makebox{$m(y)$}}
\put(123,74){\makebox{$1-\bc$}}
\put(7,22){\makebox{$\bc$}}
\put(123,8){\makebox{$m_{\rm V}$}}
\put(-10,77){\makebox{$p(y),m(y)$}}
\put(0,0){
        \makebox{
                \setl
                \epsfbox{f12.epsf}}
        }
\end{picture}
\caption{
\label{figxv}
Scaling of the shock: Density profiles 
of positive and negative particles
for $\pb=0.5$ fixed. Data for $L=$100, 200, 400 sites.
The solid curve fitted to the shock is
given by Eq.(\protect\ref{esssvi}).
Eq.(\protect\ref{esssv}) gives the expression of $k_0$.
         }
\end{figure}
}
%
%
\def\figxvii{
\begin{figure}
\def\sgr{\scriptstyle}
\setlength{\unitlength}{1mm}
\def\setl{ \setlength\epsfxsize{12cm}}
\begin{picture}(155,90)(-20,-2)
\put(114,-1){\makebox{$z$}}
\put(105,50){\makebox{$p(z)$}}
\put(35,20){\makebox{$m(z)$}}
\put(0,0){
        \makebox{
                \setl
                \epsfbox{f13.epsf}}

        }
\end{picture}
\caption{
\label{figxvii}
Density profile $p(z)$ and $m(z)$ for $\de=0.05$ and $L=200$.
         }
\end{figure}
}
%
\def\figmodel{
\setlength{\unitlength}{10pt}
\def\shoi{\begin{picture}(14,11)(-1,-3)
	\put(0,0){\vector(1,0){11.5}}
	\put(0,0){\vector(0,1){7}}
	\put(4,0){\line(0,1){0.3}}
	\put(10,0){\line(0,1){0.3}}
	\put(0,5){\line(1,0){0.3}}
	\put(-0.7,1){\makebox(0,0)[c]{$\scriptstyle \mb_{\rm V}$}}
	\put(-0.7,2){\makebox(0,0)[c]{$\scriptstyle \pb_{\rm U}$}}
	\put(-0.7,5){\makebox(0,0)[c]{$\scriptstyle \pb_{\rm V}$}}
	\put(0,-0.7){\makebox(0,0)[c]{$\scriptstyle 0$}}
	\put(4,-0.7){\makebox(0,0)[c]{$\scriptstyle g$}}
	\put(10,-0.7){\makebox(0,0)[c]{$\scriptstyle 1$}}
	\put(11,-0.5){\makebox(0,0)[c]{$\scriptstyle z$}}
	\thicklines
	\put(0,1){\line(1,0){0.25}}
	\put(0.75,1){\multiput(0,0)(1,0){9}{\line(1,0){0.5}}}
	\put(9.75,1){\line(1,0){0.25}}
	\put(0,2){\line(1,0){4}}
	\put(4,2){\line(0,1){3}}
	\put(4,5){\line(1,0){6}}
	\thinlines
	\put(5.5,-3){\makebox(0,0)[c]{(a)}}
	\end{picture}
	}
\def\shoiii{\begin{picture}(14,11)(-1,-3)
	\put(0,0){\vector(1,0){11.5}}
	\put(0,0){\vector(0,1){7}}
	\put(4,0){\line(0,1){0.3}}
	\put(10,0){\line(0,1){0.3}}
	\put(0,2){\line(1,0){0.3}}
	\put(-0.7,1){\makebox(0,0)[c]{$\scriptstyle \pb_{\rm R}$}}
	\put(-0.7,2){\makebox(0,0)[c]{$\scriptstyle \mb_{\rm S}$}}
	\put(-0.7,5){\makebox(0,0)[c]{$\scriptstyle \mb_{\rm R}$}}
	\put(0,-0.7){\makebox(0,0)[c]{$\scriptstyle 0$}}
	\put(4,-0.7){\makebox(0,0)[c]{$\scriptstyle g$}}
	\put(10,-0.7){\makebox(0,0)[c]{$\scriptstyle 1$}}
	\put(11,-0.5){\makebox(0,0)[c]{$\scriptstyle z$}}
	\thicklines
	\put(0,5){\line(1,0){0.25}}
	\put(0.75,5){\multiput(0,0)(1,0){3}{\line(1,0){0.5}}}
	\put(3.75,5){\line(1,0){0.25}}
	\put(4,2){\line(0,1){0.25}}
	\put(4,2.75){\multiput(0,0)(0,1){2}{\line(0,1){0.5}}}
	\put(4,4.75){\line(0,1){0.25}}
	\put(4,2){\line(1,0){0.25}}
	\put(4.75,2){\multiput(0,0)(1,0){5}{\line(1,0){0.5}}}
	\put(9.75,2){\line(1,0){0.25}}
	\put(0,1){\line(1,0){10}}
	\thinlines
	\put(5.5,-3){\makebox(0,0)[c]{(b)}}
	\end{picture}
	}
\def\shoii{\begin{picture}(14,11)(-1,-3)
        \put(0,0){\vector(1,0){11.5}}
        \put(0,0){\vector(0,1){7}}
        \put(10,0){\line(0,1){0.3}}
        \put(0,-0.7){\makebox(0,0)[c]{$\scriptstyle 0$}}
        \put(10,-0.7){\makebox(0,0)[c]{$\scriptstyle 1$}}
	\put(11,-0.5){\makebox(0,0)[c]{$\scriptstyle z$}}
        \thicklines
        \put(0,1.8){\line(1,0){0.25}}
        \put(0.75,1.8){\multiput(0,0)(1,0){9}{\line(1,0){0.5}}}
        \put(9.75,1.8){\line(1,0){0.25}}
        \put(0,1.2){\line(1,0){10}}
        \thinlines
        \put(5.5,-3){\makebox(0,0)[c]{(c)}}
        \end{picture}
        }
\begin{figure}
\def\sgr{\scriptstyle}
\setlength{\unitlength}{10pt}
\begin{picture}(46,10)(0,0)
\put(0,0){
        \makebox{\shoi\shoiii\shoii}
	}
\end{picture}
\caption{
\label{figmodel}
The shocks appearing at the first-order phase transition.
The solid and broken lines give the profile of positive and
negative particles, respectively.
}
\end{figure}
}
\section{Description of the first-order phase transition}

\figviii
We are now going to study in detail the point $\alpha=1$, 
$\beta=\bc=0.275$
where as explained in the last section we have seen a first-order phase
transition. 
The physics of the steady state can be understood only if one
considers the two order parameters $p$ and $m$. 
 
In Fig.\ref{figviii} we show, in the $p$-$m$-plane, 
the contour lines of the \FEF\ $f_L (p,m)$ near its minima.
The figure
suggests a large $L$ limit in which we get the following picture: 
one
line segment parallel to the $p$-axis (from U to V in Fig.\ref{figviii}), 
one line segment parallel
to the $m$-axis (from S to R), a smooth curve joining the points S, T and U.
Let us give the ($p,m$) coordinates of these five points: V=(0.725,0.149) 
where $p$ and
$m$ are obtained from the mean-field equation (\ref{eintx}), 
R=(0.149,0.725), 
obtained from
Eq.(\ref{eintx}) exchanging $p$ and $m$, 
T=(0.216,0.216), 
obtained from Eq.(\ref{eintviii}).
It is convenient to denote by $p_{\rm A}$ and $m_{\rm A}$ 
the $p$ and $m$ coordinates of the
point A. 
With this notation the coordinates of the points U and S are
U=($1-p_{\rm V},m_{\rm V}$) and S=($p_{\rm R},1-m_{\rm R}$). 
The points V and R are to be
expected to play a special role since they represent the broken phase for
a value of $\beta$ slightly smaller than $\bc$. 
Also the point T is to be
expected since it corresponds to the symmetric phase ($\beta$ slightly higher
than $\bc$). 
The points S and U are unexpected since they do not
correspond to any phase. 
Their coordinates have however a remarkable
property: $p_{\rm U}=1-p_{\rm V}$ and 
similarly $m_{\rm S}=1-m_{\rm R}$. 
These relations are typical for
shocks (see Appendix A and Ref.\cite{cxii}).
\figmodel
\figx
\figxi
\figxii
\figxv
\figxvii
The following simple model captures the physics of the phase
transition. First let us introduce the scaling variable $z=k/L$,
where $k$ is
the site variable and $L$ the length of
the lattice. 
The model will apply in the limit where $k$ and $L$ are large. 
We
assume that the following processes take place:
\begin{itemize}
\item
With a probability $P_1$ we have shocks of uncorrelated positive particles as
described by Fig.\ref{figmodel}a. 
The probability to find a positive particle at the
left of the point $g$ with a density $p_{\rm U}$ 
and to the right of the point $g$ 
with a density $p_{\rm V}$ is independent of $z$. 
One has fronts for $0<g<1$ with
equal probability (this corresponds to the U-V line segment). 
The negative
particles are also uncorrelated and 
have a constant density $m_{\rm V}$.
\item
Also with a probability $P_1$ we have shocks of negative particles (see
Fig.\ref{figmodel}b) 
which are just the $CP$-reflected shocks of the positive
particles (this corresponds to the R-S line segment).
\item
With a probability $P_2$ we have configurations (Fig.\ref{figmodel}c) 
of constant density
(independent of $z$) corresponding to all pairs of values $p$ and $m$ 
along the
line segment S-U, all with the same probability.
\end{itemize}
Obviously
\begin{equation}
2P_1+P_2=1.
\end{equation}
The assumption that we have a line segment between S and U is an approximation
since this line segment intersects the $p=m$ line in the point of 
coordinates
\mbox{($(p_{\rm U}+p_{\rm R})/2,(m_{\rm S}+m_{\rm V})/2$)}=($0.212,0.212$) 
which does not coincide with T but is very close to it. 
We
did not try to find a better description of the S-T-U curve because this
would imply many more Monte Carlo data than we were able to collect.

The probabilities $P_1$ and $P_2$ are determined 
from the condition that $f_L(p,m)$
has the same 
value everywhere on the ``boomerang''. 
We get
\begin{eqnarray}
\label{esssii}
\def\ri{m_{\rm V}}
\def\rii{p_{\rm U}}
\def\riii{p_{\rm V}}
P_1=P_2=\frac{\riii-\rii}{2(\riii-\ri)}
        =0.39
\\
\def\ri{m_{\rm V}}
\def\rii{p_{\rm U}}
\def\riii{p_{\rm V}}
P_2=\frac{\rii-\ri}{\riii-\ri}
        =0.22\;.
\end{eqnarray}

Let us now check the model, in this way we can present the data in an
organized way. 
First, we can can compute $f(d)$.
One obtains a function
which is flat between  $-(p_{\rm V}-m_{\rm V})$ and 
$(p_{\rm V}-m_{\rm V})$. 
This is just what one sees
in Fig.\ref{figiv} for $\bc$. 
Next we compute $m(z)$ and get
\begin{eqnarray}
\label{esssiii}
\fl m(z)&=P_1 m_{\rm V} + \frac 12 P_2 (m_{\rm V}+m_{\rm S}) 
	+ P_1 (m_{\rm R}-z (m_{\rm R}-m_{\rm S}))
	&=0.387-0.175\,z\; .
\end{eqnarray}
Obviously $p(z)=m(1-z)$.
In Fig.\ref{figx} we show the density of negative particles 
as a function of $z$ for
various numbers of sites. 
One notices that the straight
line given by Eq.(\ref{esssiii})
fits the data.

In order to make sure that the linear density
distribution observed in Fig.\ref{figx} is not a finite-size effect, 
in Fig.\ref{figxi} we show
for $\beta=0.26$ (below the critical point) 
the function $m(z)$. 
One observes
that the data converge (slowly) from below to the 
mean-field value $m=0.44$   
obtained taking the average of the two equations (\ref{eintx}). 

In Fig.\ref{figxii} we show
$m(z)$ for $\beta=0.3$ (above $\bc$).
The data converge from above to the
value $m=0.23$ which is obtained from Eq.(\ref{eintviii}). 
In conclusion, the non-constant
distributions seen just below and just above $\bc$ are cross-over phenomena.

In order to ``see'' the shocks, we have looked at configurations where the
density of positive particles $p$, averaged over the lattice, is fixed. 
This corresponds to a point on
the U-V line segment in Fig.\ref{figviii} and one expects 
to see a shock like in Fig.\ref{figmodel}a.
Since we are also interested to see what kind of finite-size effects exist
(see Appendix A), we have chosen to present the densities as functions not
of $z$ but in terms of the variable
\begin{equation}
\label{esssiv}
y=\frac{k-k_0}{\sqrt{L}}
\end{equation}
where $k_0$ is the position of the front which in the large $L$ limit
is given by
\begin{equation}
\label{esssv}
k_0=\frac{L}{2\bc-1}(\pb-1+\bc).
\end{equation}
The data for $p=0.5$ are shown in Fig.\ref{figxv}. 
Let us leave aside for a moment the
existence of ``tails'' on the left hand side of the figure. 
The data scale,
they are independent of $L$. 
A good fit to the data is given by 
\begin{equation}
\label{esssvi}
p(y)=(\frac12-\bc) \erf (y/2)+\frac12
\end{equation}
Obviously, if we show the data as a function of $z$ one gets the front
from Fig.\ref{figmodel}a with
$g=0.5.$ 
The value of $m$ in Fig.\ref{figxv} 
corresponds to the value of $m_{\rm V}$ in Fig.\ref{figmodel}a as expected.
We now turn our attention to the ``tails''. 
For the positive particles one
can fit the data like
\begin{equation}
\label{esssviii}
p(k)=\mu \,k^{-\frac12} + \bc
\end{equation}
where $k$ is the position of the site starting with the left hand side of
the chain and $\mu=0.08$ is a constant. 
We have checked that for other shocks (one
takes $p$ different of $0.5$ but still on the U-V segment of Fig.\ref{figviii}) 
the value
of $\mu$ stays unchanged. 
For negative particles the tails are exponential
with a correlation length independent of $L$. 
This observation is intriguing
because common sense would suggest that in this case
one should have a finite correlation
length also in the time direction, i.e. masses in the spectrum of the problem
and not only the massless excitations discussed in Sec.2.

Another test of the model, this time on a part of it on which we certainly
expect to be less precise, is to ``cut'' through the handle of the boomerang.
We have taken $d=0.05$ in order to check if profiles like in 
Fig.\ref{figmodel}c 
are seen.
The data are shown in Fig.\ref{figxvii} 
together with the expected constant values
(take the intersection of the line $d=0.05$ with the line segment S-U 
in Fig.\ref{figviii}).
As one notices we find agreement although some mini-shocks are not excluded.
After presenting the data we are left with a puzzle. We have expected the
point T to be singled out in some way (it represents the symmetric phase),
but it is not.
\def\tabii{
\begin{table}
\caption{The minima $\de_L^{\rm min}$
of $f_L(\de)$ (error $\pm 0.01$) 
and the mean-field predictions $\de_{\rm m-f}$ for 
$\alpha=1$, various $\beta$ and $h$.
Values of $\de_L^{\rm min}$ predicted (within $10\%$) are printed bold face.
Several values given for a set of parameters $\alpha$, $\beta$, $h$ 
reflect different minima ($\de_L^{\rm min}$) or 
different stationary solutions of the mean-field 
equations ($\de_{\rm m-f}$).}
\begin{indented}
\item[]
\begin{tabular}{@{}llllllllllllllllllllllll}
\br
$\beta$ & $h$ & $L$ & $\de_L^{\rm min}$& $\de_{\rm m-f}$
\\
\mr
0.6	&0	&200	&{\bf 0.00}	&0\\
	&0.1	&200	&{\bf -0.08}	&-0.0943\\
\mr
0.35    &0	&400	&{\bf 0.00}	&0\\
	&0.02   &100	&-0.05 		&-0.468\\
	&0.1	&100	&-0.45		&-0.512\\
\mr
0.329	&0	&400	&0.00		&-0.490/+0.490/0\\
	&0.02	&200	&-0.06		&-0.501\\
	&0.1	&200    &{\bf -0.52}	&-0.543\\
\mr
0.3	&0	&400	&0.00		&-0.537/+0.537/0\\
	&0.02	&400	&-0.15		&-0.546/+0.527/+0.098\\
	&0.1	&400	&{\bf -0.58}	&-0.584\\
\mr
0.26	&0	&200	&{\bf -0.53}/{\bf +0.53}	&-0.600/+0.600/0\\
	&0.01	&100	&{\bf -0.53}/+0.43	&-0.604/+0.596/+0.019\\
	&0.02	&100	&{\bf -0.54}		&-0.607/+0.592/+0.038\\
	&0.07	&100	&{\bf -0.61}		&-0.629\\
\mr
0.1	&0	&200	&{\bf -0.84}/{\bf +0.84}	&-0.849/+0.849/0\\
	&0.02	&100	&{\bf -0.84}/{\bf +0.84}	&-0.852/+0.846/+0.00519\\
	&0.05	&100	&{\bf -0.85}/{\bf +0.83}	&-0.856/+0.841/+0.013\\
	&0.1	&50	&{\bf -0.86}/{\bf +0.81}	&-0.864/+0.833/+0.026\\
\br
\end{tabular}
\end{indented}
\end{table}
}
\def\figxviii{
%
\begin{figure}[tb]
\def\sgr{\scriptstyle}
\setlength{\unitlength}{1mm}
\def\setl{ \setlength\epsfxsize{12cm}}
\begin{picture}(155,90)(-20,-2)
\put(114,-1){\makebox{$\de$}}
\put(-5,77){\makebox{$f_L(\de,h)$}}
\put(100,20){\makebox{$h=0$}}
\put(100,38){\makebox{$h<\hc$}}
\put(100,58){\makebox{$h=\hc$}}
\put(100,73){\makebox{$h>\hc$}}
\put(0,0){
        \makebox{
                \setl
                \epsfbox{f15.epsf}}
        }
\end{picture}
\caption{
\label{figxviii}
$f_L(\de,h)$ for $L=30$, $\alpha=1.0$, $\beta=0.1$ and
$h=0,\, 0.1,\, 0.2281,\, 0.3$.
For $h=0$ the system is in the phase $\tilde{\rm D}$. 
         }
\end{figure}
}
\def\figxxx{
\begin{figure}
\def\sgr{\scriptstyle}
\setlength{\unitlength}{1mm}
\def\setl{ \setlength\epsfxsize{12cm}}
\begin{picture}(155,90)(-20,-2)
\put(114,-1){\makebox{$\alpha$}}
\put(4,77){\makebox{$\beta$}}
\put(100,70){\makebox{A}}
\put(50,40){\makebox{$\tilde{\rm B}$}}
\put(100,15){\makebox{$\tilde{\rm D}$}}
\put(0,0){
        \makebox{
                \setl
                \epsfbox{f14.epsf}}
        }
\end{picture}
\caption{
\label{figxxx}
Phase diagram of the two-species model
for $h=0$ obtained from the Monte Carlo simulations.
The phase A is given by mean field.
         }
\end{figure}
}
\def\figxix{
\begin{figure}[tb]
\def\sgr{\scriptstyle}
\setlength{\unitlength}{1mm}
\def\setl{ \setlength\epsfxsize{12cm}}
\begin{picture}(155,90)(-20,-2)
\put(114,-1){\makebox{$\de$}}
\put(0,77){\makebox{$f_L$}}
\put(0,0){
        \makebox{
                \setl
                \epsfbox{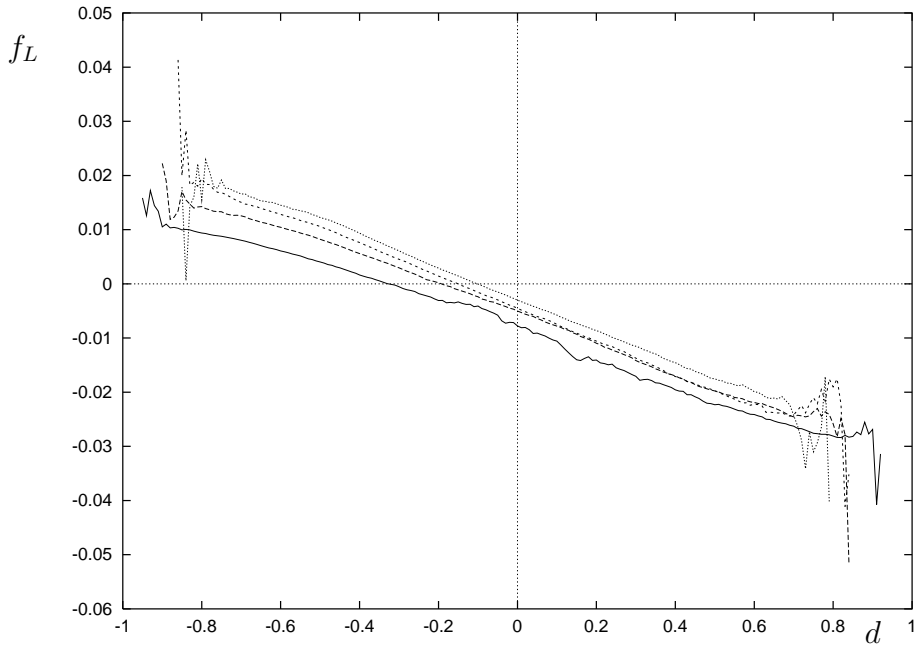}}
        }
\end{picture}
\caption{
\label{figxix}
$f_L(\de,h=0)-f_L(\de,h=0.02)$ for
$\beta=$0.20, 0.26, 0.28, 0.30 and $L=100$ sites.
         }
\end{figure}
}
\def\figxxi{
%
\begin{figure}[tb]
\def\sgr{\scriptstyle}
\setlength{\unitlength}{1mm}
\def\setl{ \setlength\epsfxsize{12cm}}
\begin{picture}(155,90)(-20,-2)
\put(114,-1){\makebox{$L$}}
\put(0,77){\makebox{$T$}}
\put(100,28){\makebox{$\Ts$}}
\put(100,60){\makebox{$\Tl$}}
\put(0,0){
        \makebox{
                \setl
                \epsfbox{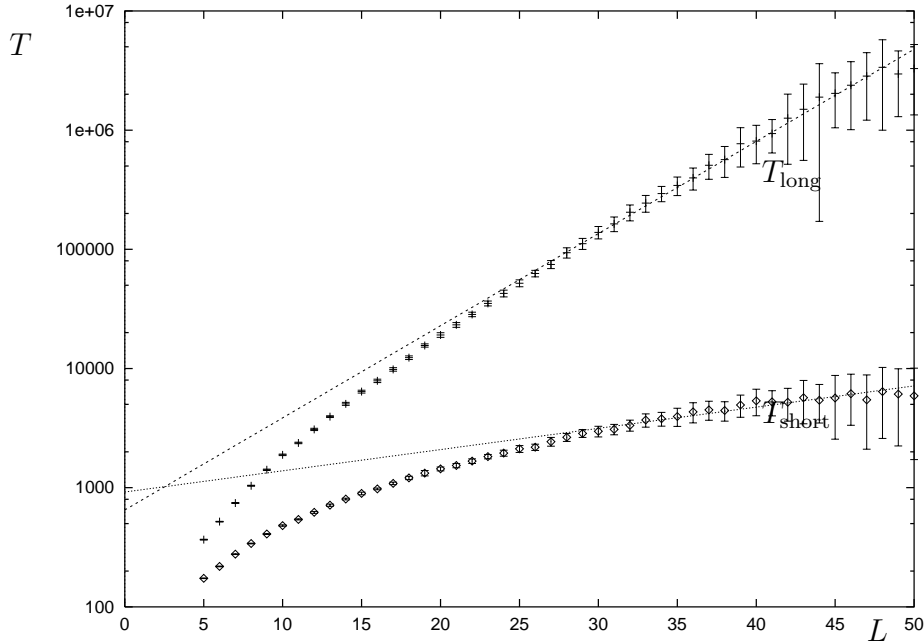}}
        }
\end{picture}
\caption{
\label{figxxi}
Logarithmic plot of the flip times for $\alpha=1.0$, $\beta=0.1$ and $h=0.1$.
The dotted curves are given by (\protect\ref{essssiii}).
         }
\end{figure}
}
\section{The broken phase in the presence of a symmetry 
breaking external source}
\figxxx
\figxviii
\figxix
\tabii
\figxxi
As explained in Sec.2 and shown in Fig.\ref{figxxx},
the phase diagram of the $CP$-symmetric model 
consists
of a power law phase A and a broken phase $\tilde{\mbox{D}}$ 
separated by a first-order phase transition line
from the unbroken phase $\tilde{\mbox{B}}$. 
In
equilibrium systems, 
if one has a transition from an ordered to a disorderd
phase, the broken phase corresponds to first-order phase
transitions (and the \FEF\ is a flat function between 
the two phases). 
As we
have seen this is not the case for steady states and for this reason we
use the expression ``first-order'' only on the separation line where the 
\FEF\ is
flat and the expression ``broken phase'' for the domain where the 
\FEF\ is not
convex and has two minima.

We consider now the effect of an explicit $CP$ symmetry breaking in the
model. We take $h>0$ in the boundary rates given by equations 
(\ref{eintii}) and (\ref{eintiii}).
Using Monte Carlo simulations, we have computed the 
\FEF\ $f_L(\de,h)$ taking $\alpha=1$
and several values of $\beta$ and $h$; 
$\de$ is again the difference between the average
densities of positive and negative particles. In Fig.\ref{figxviii}
we show $f_L(\de,h)$
for $\beta=0.1$ (in the $\tilde{\mbox{D}}$ phase) 
and several values of $h$. We observe that below
$\hc=0.2281$ the \FEF\ shows two minima and only one minimum for 
$h>\hc$.
For $h=\hc$ the first and second derivative of $f_L(\de,h)$ 
with respect to $\de$ vanishes.
This is
the definition of a spinodal point
\cite{cxvi}.

The behavior of $f_L(\de,h)$ is the one
expected from a Ginzburg-Landau ansatz \cite{GiLa}. 
Our data are of course not
precise enough to give an estimate with four digits for $\hc$, 
the value we
give is obtained in mean-field. 
We will return soon to the problem of the
validity of mean-field. 

In the Ginzburg-Landau ansatz $f(\de,h)$ 
has a simple $h$
dependence:
\begin{equation}
\label{essssi}
f(\de,h)=f(\de,h=0)+\nu\, h\, \de +g(h)
\end{equation}
where $\nu$ is a parameter. We have collected data for various values 
of $\beta$ and $h$
and find $\nu= 1.4$ getting a fit to the data within $10\%$. 
In Fig.\ref{figxix} we fix
the value of $h =0.02$ and show the difference
\begin{equation}
\label{essssii}
f_L(\de,h=0)-f_L(\de,h=0.02)
\end{equation}
as a function of $\de$. According to Eq.(\ref{essssi}) 
we expect straight lines with
the slope $-0.028$. 
The data are compatible with this prediction.
We have repeated the simulations for several values
of $h$ with the same result.

We have computed the mean-field approximation of the model
(\ref{einti}--\ref{eintiii}) for $h$
different from zero \cite{cxxix}, 
extending the calculation done in Ref.\cite{cvii}. 
Our
motivation was to find out if, like in the $h=0$ case, 
one can understand in a
simple way where and in which way mean-field is exact or wrong. 
In Table 2, for
different values of $\beta$ and $h$, we give the values of $\de_L^{\rm min}$ 
which
denotes the minima of the \FEF\ as determined by 
Monte Carlo simulations
as well as the values of $\de_{\rm m-f}$ 
obtained in mean-field. Comparing the
Monte-Carlo simulations with the mean-field results we reach the conclusion
that probably mean-field is not exact anywhere.
It is however a good
approximation for small values of $\beta$ and large values of $h$.
 
We are now going to discuss some aspects of the dynamics of the model when
one includes the symmetry breaking term. Our attention goes first to the
spinodal point and we look at the spectrum of the hamiltonian 
(see Appendix B). 
We notice that the first excited state has an energy which vanishes in
the large $L$ limit. The dynamical critical exponent 
(see Eq.(\ref{essiii})) determined
using the
values of the first excited state of chains up to 11 sites 
$\beta=0.1$ and $\hc=0.22810$ as given by mean-field
is
$z= 1.00\pm 0.01$. 
We have repeated the same calculation for $\beta=0.05$ ($\hc=0.28096$)
and found 
$z=1.000\pm0.001$. 
As discussed previously, it is not clear if the values of
$\hc$ obtained by mean-field can be trusted to a precision of
five digits. To our knowledge 
it is for the first time that the exponent $z=1$
appears in stochastic processes. 

In Ref.\cite{ArHe} 
the toy-model of
Godr{\`e}che \etal \cite{GoLuEvMuSa} is examined 
(which corresponds to the $\beta\rightarrow 0$ limit of
the present model) and again $z=1$ is found. 
More than that, in Ref.\cite{ArHe} it is shown
that the spectrum has massless excitations 
which are universal 
and also
massive excitations. (As 
opposed to the present model where we could study properly only one
level, in the toy-model we can study many of them.)
The toy-model has two parameters, one
which corresponds to $\alpha$ in the original model and $h$. 
Normalizing the
hamiltonian such that the sound velocity is
the same, at the spinodal point
the spectra corresponding to the massless regime do not depend
on the former parameter.    
One can ask which implications has a $z=1$ exponent. 
In Ref.\cite{ArHe} several
properties of the time dependence of the order parameter are given. The
simplest one being that the flip time \cite{cvi} from a configuration 
with the $d$ value of
the spinodal point to a configuration corresponding to the minimum
of $f_L(\de,h)$ (see Fig.\ref{figxviii}) should be proportional to $L$.

We would like to make another observation on the flip times related to
barriers in the \FEF . 
Let us take $\beta=0.1$ and $h=0.1$ 
in Fig.15 and denote by
$\Delta_1$ ($\Delta_2$) the difference between the maximum of 
the \FEF\ and the minimum
of the \FEF\ 
for negative (positive) values of $\de$
($\Delta_2=0$ at the spinodal point). From 
the Monte Carlo data ($L=50$ sites)
we get $\Delta_1=0.178\pm0.002$ and $\Delta_2=0.041\pm0.002$. 
The two flip times
from the stable (unstable) to the unstable (stable) 
phase are
$\Tl$ ($\Ts$). 
They have been measured for chains of different lengths $L$
and their values are shown in Fig.\ref{figxxi}. 
The straight lines in Fig.\ref{figxxi} 
are given by the ansatz
\begin{eqnarray}
\label{essssiii}
\Tl \propto \exp (\Delta_1 \;L)
\nonumber\\
\Ts \propto \exp (\Delta_2 \;L) \;.
\end{eqnarray}
The agreement between the ansatz (\ref{essssiii}) 
and the data shows a connection
between the dynamics of the model (flip times) and the properties of the
steady state (the \FEF). 
Repeating the same procedure for a different
value of $\beta$ gives the same result. 
More about this subject can be found in
Ref.\cite{ArHe}.
\section{Conclusions}

We have shown in the framework of the two-species model the following
pattern of spontaneous $CP$ symmetry breaking for steady states. 
If we start in
the ``disordered'' phase 
(where the free energy functional  as a function of
the order parameter has only one minimum) and lower the ``temperature'' 
(the
parameter $\beta$) we have a first-order phase transition at $\bc$ with a
flat free energy functional (like in equilibrium problems). 
At $\bc$ the
density profiles and correlation functions in the scaling regime can be
explained using a certain combination of shocks.
Below the critical ``temperature'' $\bc$, in the broken phase, the free
energy functional is not a convex function of the order parameter (unlike
in equilibrium problems) but has two minima corresponding to two phases
defined by their average densities of positive and negative particles.
If the $CP$ symmetry of the model is  broken by a small perturbation
(parameter $h$),
one has only one minimum for the first-order phase transition
(like in equilibrium) but still two minima in the broken phase. If we
increase $h$ one gets spinodal points.

At the first-order phase transition we have seen massless excitations in
the spectrum of the quantum chain hamiltonian associated with the dynamics
of the process with a critical dynamical exponent $z=2$. 
This does not exclude
the presence of massive excitations which are suggested by the exponential
tails seen in Fig.\ref{figx} and Fig.\ref{figxv} in the density profiles. 
Massive excitations
are not seen for the one-species model described in Appendix A \cite{ULI}.

At the spinodal points we have also seen massless excitations 
with $z=1$ as
well as massive excitations. The latter are suggested by exponential tails
observed in the density profiles not shown in this paper. These
observations are confirmed by the investigation of the low $\beta$ limit of
the model \cite{ArHe}.
At this point it is not clear if the exponent $z=1$ has profound implications  
like in equilibrium problems where it implies conformal invariance or is
just one dynamical critical exponent among others. We have looked to
conformal towers in the spectrum of the hamiltonian but the data of our
short quantum chains did not give any clean results. This is probably due
to the fact that the critical point is not known exactly (we used the
values obtained from mean-field) and that in general the convergence to the
thermodynamic limits is slower in stochastic quantities than in
equilibrium ones. This is an empirical observation.
We think that a further investigation of the model is necessary by taking
the second rate in (\ref{einti}) not equal to the other two. Some preliminary
results are described in Ref.\cite{cvii}. 
This study can be done not only by Monte
Carlo simulations but also using the analytical methods developed recently in
Ref.\cite{cxxxi}.

\ack
We would like to thank M.~Evans, C.~Godr{\`e}che, J.~M.~Luck, D.~Mukamel
and especially B.~Derrida for discussions. 
We would like also to thank SISSA
for hospitality.
\section{The asymmetric exclusion process}
\def\figai{
\begin{figure}[th]
\def\sgr{\scriptstyle}
\setlength{\unitlength}{1mm}
\def\setl{ \setlength\epsfxsize{12cm}}
\begin{picture}(155,90)(-20,-2)
\put(114,-1){\makebox{$c$}}
\put(0,77){\makebox{$f(c,h)$}}
\put(0,0){
        \makebox{
                \setl
                \epsfbox{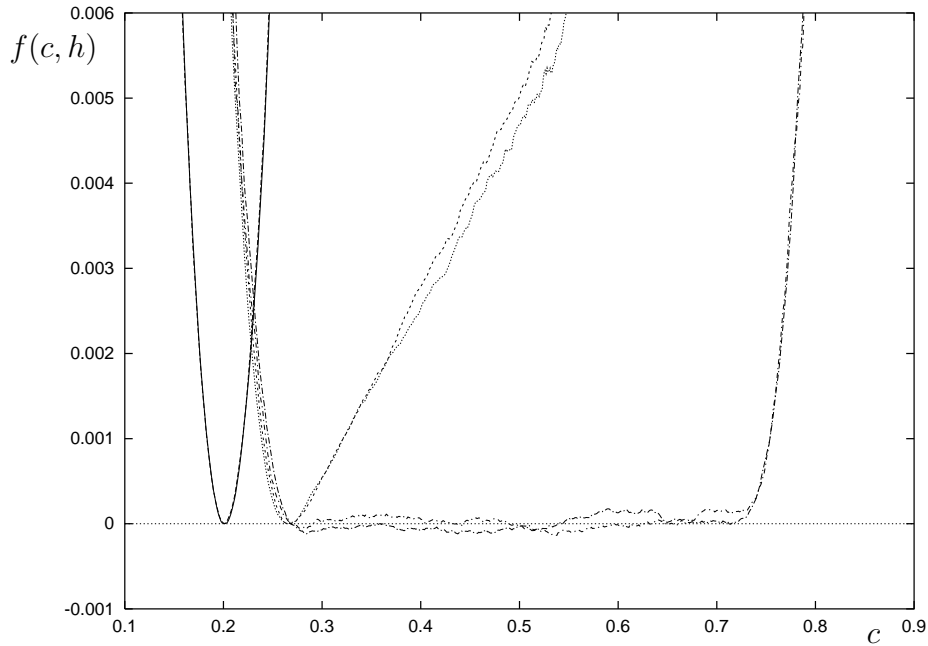}}
        }
\end{picture}
\caption{Asymmetric diffusion model: $f_L(c,h)$
	for $s=0.25$ and $h=0$, 0.004, 0.2, each for
	$L=800$ and 1000 sites.
	We subtracted $f_L(c_{\rm min},h)$ from each curve.
\label{figai}
         }
\end{figure}
}
\def\figaii{
\begin{figure}[th]
\def\sgr{\scriptstyle}
\setlength{\unitlength}{1mm}
\def\setl{ \setlength\epsfxsize{12cm}}
\begin{picture}(155,90)(-20,-2)
\put(114,-1){\makebox{$y$}}
\put(0,77){\makebox{$c(y)$}}
\put(0,0){
        \makebox{
                \setl
                \epsfbox{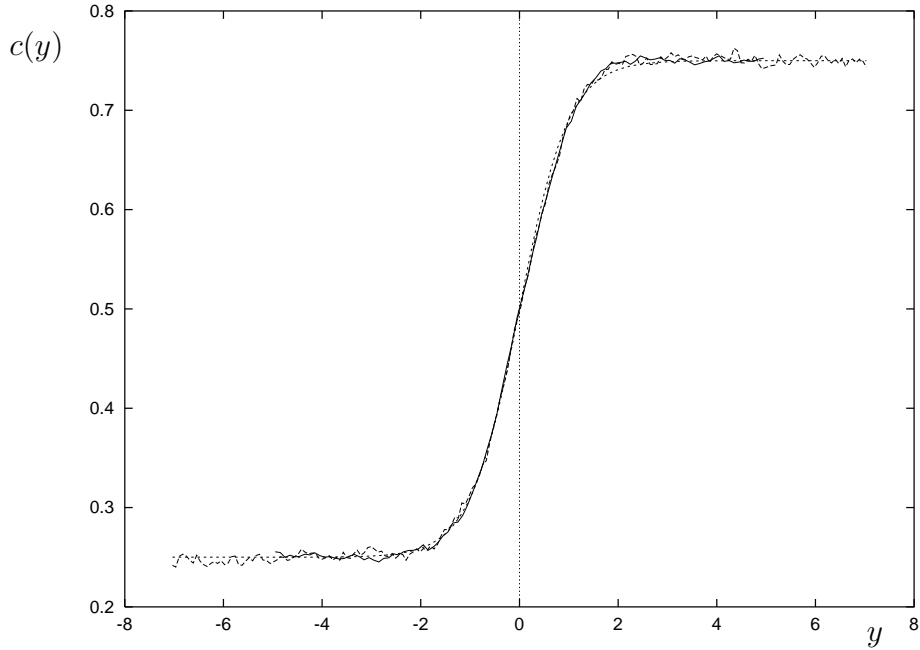}}
        }
\end{picture}
\caption{
\label{figaii}
Scaling of the shock: Density profiles 
for $\alpha=\beta=0.25$
(with $c=0.5$ fixed) and for $L=$100, 200.
The dotted curve is $0.25\tanh (y)+0.5$.
The wriggled curves represent the Monte Carlo data which
coincide with the solid curve within $1\%$.
         }
\end{figure}
}
The physics of this model is well understood \cite{cii,ciii,civ}, 
here we are going to
discuss some aspects of it which are necessary 
for the understanding of the finite-size effects of
the two-species model.
We consider a chain of length $L$ with two types of particles ($+$ and $-$). 
In an
infinitesimal time step $\d t$ the following processes take place:
\begin{eqnarray}
(+)_k\,(-)_{k+1}\,&\rightarrow (-)_k\,(+)_{k+1}
\quad&\mbox{with probability } \d t
\nonumber
\\
(-)_1 &\rightarrow (+)_1
\,\quad&\mbox{with probability }\alpha\; \d t
\\
(+)_L &\rightarrow (-)_L
\quad&\mbox{with probability }\beta \;\d t.
\nonumber
\end{eqnarray}
\def\s{\sigma}

Defining $\s=\alpha+\beta$ 
and $h=\beta-\alpha$ one notices that for $h=0$ the model is $CP$-invariant. 
We denote by $c$ the average density of positive particles.
The model presents three phases:
\begin{itemize}
\item
the power law phase ($\alpha> 1/2, \beta >1/2$) where $c=1/2$
\item
the low density phase ($\alpha<1/2, \beta >\alpha$) where $c=\alpha$
\item
the high density phase ($\alpha>\beta, \beta<1/2$ ) where $c=1-\beta$
\end{itemize}
A coexistence line ($\alpha=\beta <1/2$) 
separates the low density from the high
density phase.
The concentration has a discontinuity along this line from
$c=\beta$ to $c=1-\beta$.

\figai
\figaii
Using Monte Carlo simulations
we have computed the \FEF\ $f_L(c,h)$ 
keeping $\s$
fixed and using several values of $h$. 
The results are shown in Fig.\ref{figai}. 
We notice
that for $h=0$, the \FEF\ is a flat function between $c=\beta$ 
and $c=1-\beta$. 
A small
positive $h$ ($h=0.004$) gives a \FEF\ 
with only one minimum at $c=\beta$ just like in
equilibrium first-order phase transitions. 
A larger value of $h$ ($h=0.1$)
moves the minimum away. 
We have studied the convergence of the function 
$f_L(c,h)$ to $f(c,h)$ taking
$c_{\rm min}$ the value of $c$ 
where the \FEF\ has a minimum. 
For all the examples we
have studied we have found that
\begin{equation}
f_L(c_{\rm min},h)= f(c_{\rm min},h) +a/L
\end{equation}
Where the constant $a$ depends on $\s$ and $h$. 
For example if we take $\s=0.5$ and
$h=0.2$ like in Fig.\ref{figai} we find $a=4.2\pm 0.2$.
We have also found that if we start on the coexistence line with a certain
value of $\s$ and switch-on the $CP$ symmetry breaking term $h$, 
we have the
following simple expression for the \FEF :
\begin{equation}
f(c,h)=f(c,h=0)+ \nu\, c\, h +g(h)
\end{equation}
where $\nu$ depends on $\s$. 
For the example $\s=0.5$ studied above we find $\nu=4.5$.
We have observed that if we take $L=200$ the \FEF\
has already a shape independent of $L$, the single $L$
dependence being in $f_L(c_{\rm min},h)$.

We would like to mention that much of the physics which will be described
below is contained in a simplified model studied by Sch{\"u}tz \cite{cxxv}. 
In this model
the bulk dynamics is deterministic and only the boundaries are stochastic.
The critical dynamical exponent on the coexistence line 
of the present model
is $z=2$ \cite{ULI} 
like in Ref.\cite{cxxv}.
This corresponds to the picture in which the front 
of a shock performs a symmetric random 
walk.

On the coexistence line the density of positive particles is high at the right
hand side of the chain ($c=1-\beta$) and low at 
the left hand side of the chain
($c=\beta$, $\beta<1/2$). 
The two regimes are joined by a shock front. One
measures such a shock by constraining the average density of positive
particles to a given value between $\beta$ and $1-\beta$. 
If $c(k)$ is the concentration
at lattice site $k$, it is useful to define the variable
\begin{equation}
y=\frac{k-k_0}{\sqrt{L}}
\end{equation}
where
\begin{equation}
k_0=\frac{L}{2\beta-1}(c-1+\beta)
\end{equation}
is the position of the shock front.
We have found empirically that the Monte Carlo data are fitted 
(within $1\%$)
by the function
\begin{equation}
c(y)=(\frac12-\beta) \tanh (y)+\frac12\; .
\end{equation}
In Fig.\ref{figaii} we show the Monte Carlo data for $c=0.5$ 
and $\alpha=\beta=0.25$.
The existence of 
a front with width of order $L^{1/2}$ is to be expected
from the model studied by Sch{\"u}tz \cite{cxxv}, 
from the exclusion model with a blockage \cite{JaLe,cxxvi}
and from the asymmetric model with an impurity \cite{cxxvii}.
Our purpose in studying the finite-size behavior of the shocks was to
find out how many lattice sites are necessary to be in the scaling limit
in order to know for the three-states model which lattice size are
necessary to see shocks. It turns out that $L=100$ are enough.
\section{The Hamilton operator}
The time evolution operator $H$ of the master equation (\ref{eintiv})
for the processes (\ref{einti})--(\ref{eintiii})
can be written as a sum of operators acting non-trivially
on adjacent sites only (hopping)
and of two operators acting on the first and last site
(input and output)\cite{cix}:
\begin{equation}
H= H_1 + \sum_{k=1}^{L-1} H_{k,k+1} + H_L
\label{ham}
\end{equation}
The matrices are
\begin{eqnarray}
\fl
H_{k,k+1}=I_1\otimes\cdots\otimes I_{k-1} \otimes
		\left( \begin{array}{rrrrrrrrr}
		0& 0& 0& 0& 0& 0& 0& 0& 0\\
		0& 0& 0&-1& 0& 0& 0& 0& 0\\
		0& 0& 1& 0& 0& 0& 0& 0& 0\\
		0& 0& 0& 1& 0& 0& 0& 0& 0\\
		0& 0& 0& 0& 0& 0& 0& 0& 0\\
		0& 0& 0& 0& 0& 1& 0& 0& 0\\
		0& 0&-1& 0& 0& 0& 0& 0& 0\\
		0& 0& 0& 0& 0&-1& 0& 0& 0\\
		0& 0& 0& 0& 0& 0& 0& 0& 0
		\end{array}
	\right)_{k,k+1}\!\!\!\!\!\!\!\!\!
	\otimes I_{k+2}\otimes\cdots\otimes I_L\nonumber\\[0.3cm]
\fl
H_1=\left(\begin{array}{rrr}
	\alpha& 0& -\beta(1-h)\\
       -\alpha& 0& 0\\
	     0& 0& \beta(1-h)\\
	\end{array}\right)_1\!
	\otimes I_{2}\otimes\cdots\otimes I_L\nonumber\\[0.3cm]
\fl
H_L=I_1\otimes\cdots\otimes I_{L-1} \otimes
	\left(\begin{array}{rrr}
	\alpha& -\beta(1+h)& 0\\
	     0&  \beta(1+h)& 0\\
       -\alpha&           0& 0\\
	\end{array}\right)_L          \nonumber
\end{eqnarray}
where the subscripts denotes the sites the matrices act on
and $I$ is the $3\times 3$ unit matrix.
The order of the basis at each site is chosen as 
$|0>$, $|+>$, $|->$. 
Note that $H$ is not hermitian.
\np

%
%
\end{document}